\documentclass[longauth]{aa}

\usepackage{graphicx}
\usepackage{natbib}
\usepackage{scalerel}
\usepackage{float}

\usepackage[table]{xcolor}

\usepackage{txfonts}
\usepackage[pdfencoding=auto,psdextra]{hyperref}
\hypersetup{
    colorlinks=true,
    linkcolor=blue,
    filecolor=magenta,      
    urlcolor=blue,
    citecolor=blue
}
\makeatletter
\renewcommand*\aa@pageof{, page \thepage{} of \pageref*{LastPage}}
\makeatother

\usepackage[utf8]{inputenc}

\usepackage[switch, modulo]{lineno}

\usepackage{euclid}

\usepackage[para,online,flushleft]{threeparttable}
\usepackage{orcidlink}

\graphicspath{{./}{figures/}}
\usepackage{subcaption}
\usepackage{threeparttable} 

\usepackage{subcaption} 
\usepackage[font=small]{caption} 
\usepackage[labelfont=bf]{caption} 

\newcommand{\YJHE}{\ensuremath{Y_\sfont{E},J_\sfont{E},H_\sfont{E}}\xspace}

\begin{document}
%
%

\title{\Euclid: Early Release Observations -- NISP-only sources \\
and the search for luminous $z=6$\,--\,8 galaxies\thanks{This paper is published on behalf of the Euclid Consortium}}

\newcommand{\orcid}[1]{} 
\author{J.~R.~Weaver\orcid{0000-0003-1614-196X}\thanks{\email{jweaver@astro.umass.edu}}\inst{\ref{aff1}}
\and S.~Taamoli\orcid{0000-0003-0749-4667}\inst{\ref{aff2}}
\and C.~J.~R.~McPartland\orcid{0000-0003-0639-025X}\inst{\ref{aff3},\ref{aff4}}
\and L.~Zalesky\orcid{0000-0001-5680-2326}\inst{\ref{aff5}}
\and N.~Allen\orcid{0000-0001-9610-7950}\inst{\ref{aff6}}
\and S.~Toft\orcid{0000-0003-3631-7176}\inst{\ref{aff6},\ref{aff4}}
\and D.~B.~Sanders\orcid{0000-0002-1233-9998}\inst{\ref{aff5}}
\and H.~Atek\orcid{0000-0002-7570-0824}\inst{\ref{aff7}}
\and R.~A.~A.~Bowler\orcid{0000-0003-3917-1678}\inst{\ref{aff8}}
\and D.~Stern\orcid{0000-0003-2686-9241}\inst{\ref{aff9}}
\and C.~J.~Conselice\orcid{0000-0003-1949-7638}\inst{\ref{aff8}}
\and B.~Mobasher\orcid{0000-0001-5846-4404}\inst{\ref{aff2}}
\and I.~Szapudi\orcid{0000-0003-2274-0301}\inst{\ref{aff5}}
\and P.~R.~M.~Eisenhardt\inst{\ref{aff9}}
\and G.~Murphree\orcid{0009-0007-7266-8914}\inst{\ref{aff5}}
\and I.~Valdes\orcid{0009-0002-8551-9372}\inst{\ref{aff5}}
\and K.~Ito\orcid{0000-0002-9453-0381}\inst{\ref{aff10}}
\and S.~Belladitta\orcid{0000-0003-4747-4484}\inst{\ref{aff11},\ref{aff12}}
\and P.~A.~Oesch\orcid{0000-0001-5851-6649}\inst{\ref{aff13},\ref{aff4},\ref{aff6}}
\and S.~Serjeant\orcid{0000-0002-0517-7943}\inst{\ref{aff14}}
\and D.~J.~Mortlock\orcid{0000-0002-0041-3783}\inst{\ref{aff15},\ref{aff16}}
\and N.~A.~Hatch\orcid{0000-0001-5600-0534}\inst{\ref{aff17}}
\and M.~Kluge\orcid{0000-0002-9618-2552}\inst{\ref{aff18}}
\and B.~Milvang-Jensen\orcid{0000-0002-2281-2785}\inst{\ref{aff6},\ref{aff4},\ref{aff3}}
\and G.~Rodighiero\orcid{0000-0002-9415-2296}\inst{\ref{aff19},\ref{aff20}}
\and E.~Ba\~nados\orcid{0000-0002-2931-7824}\inst{\ref{aff11}}
\and J.~M.~Diego\orcid{0000-0001-9065-3926}\inst{\ref{aff21}}
\and R.~Gavazzi\orcid{0000-0002-5540-6935}\inst{\ref{aff22},\ref{aff7}}
\and G.~Congedo\orcid{0000-0003-2508-0046}\inst{\ref{aff23}}
\and M.~Shuntov\orcid{0000-0002-7087-0701}\inst{\ref{aff24},\ref{aff3},\ref{aff4}}
\and H.~Dole\orcid{0000-0002-9767-3839}\inst{\ref{aff25}}
\and P.-F.~Rocci\inst{\ref{aff25}}
\and T.~Saifollahi\orcid{0000-0002-9554-7660}\inst{\ref{aff26},\ref{aff27}}
\and M.~Miluzio\inst{\ref{aff28},\ref{aff29}}
\and M.~Ezziati\orcid{0009-0003-6065-1585}\inst{\ref{aff22}}
\and A.~C.~N.~Hughes\orcid{0000-0001-9294-3089}\inst{\ref{aff15}}
\and J.-C.~Cuillandre\orcid{0000-0002-3263-8645}\inst{\ref{aff30}}
\and R.~Laureijs\inst{\ref{aff31}}
\and S.~Paltani\orcid{0000-0002-8108-9179}\inst{\ref{aff13}}
\and M.~Schirmer\orcid{0000-0003-2568-9994}\inst{\ref{aff11}}
\and C.~Stone\orcid{0000-0002-9086-6398}\inst{\ref{aff32}}
\and N.~Aghanim\orcid{0000-0002-6688-8992}\inst{\ref{aff25}}
\and B.~Altieri\orcid{0000-0003-3936-0284}\inst{\ref{aff28}}
\and A.~Amara\inst{\ref{aff33}}
\and S.~Andreon\orcid{0000-0002-2041-8784}\inst{\ref{aff34}}
\and N.~Auricchio\orcid{0000-0003-4444-8651}\inst{\ref{aff12}}
\and M.~Baldi\orcid{0000-0003-4145-1943}\inst{\ref{aff35},\ref{aff12},\ref{aff36}}
\and A.~Balestra\orcid{0000-0002-6967-261X}\inst{\ref{aff20}}
\and S.~Bardelli\orcid{0000-0002-8900-0298}\inst{\ref{aff12}}
\and R.~Bender\orcid{0000-0001-7179-0626}\inst{\ref{aff18},\ref{aff37}}
\and C.~Bodendorf\inst{\ref{aff18}}
\and D.~Bonino\orcid{0000-0002-3336-9977}\inst{\ref{aff38}}
\and E.~Branchini\orcid{0000-0002-0808-6908}\inst{\ref{aff39},\ref{aff40},\ref{aff34}}
\and M.~Brescia\orcid{0000-0001-9506-5680}\inst{\ref{aff41},\ref{aff42},\ref{aff43}}
\and J.~Brinchmann\orcid{0000-0003-4359-8797}\inst{\ref{aff44},\ref{aff45}}
\and S.~Camera\orcid{0000-0003-3399-3574}\inst{\ref{aff46},\ref{aff47},\ref{aff38}}
\and V.~Capobianco\orcid{0000-0002-3309-7692}\inst{\ref{aff38}}
\and C.~Carbone\orcid{0000-0003-0125-3563}\inst{\ref{aff48}}
\and V.~F.~Cardone\inst{\ref{aff49},\ref{aff50}}
\and J.~Carretero\orcid{0000-0002-3130-0204}\inst{\ref{aff51},\ref{aff52}}
\and S.~Casas\orcid{0000-0002-4751-5138}\inst{\ref{aff53}}
\and F.~J.~Castander\orcid{0000-0001-7316-4573}\inst{\ref{aff54},\ref{aff55}}
\and M.~Castellano\orcid{0000-0001-9875-8263}\inst{\ref{aff49}}
\and S.~Cavuoti\orcid{0000-0002-3787-4196}\inst{\ref{aff42},\ref{aff43}}
\and A.~Cimatti\inst{\ref{aff56}}
\and L.~Conversi\orcid{0000-0002-6710-8476}\inst{\ref{aff57},\ref{aff28}}
\and Y.~Copin\orcid{0000-0002-5317-7518}\inst{\ref{aff58}}
\and L.~Corcione\orcid{0000-0002-6497-5881}\inst{\ref{aff38}}
\and F.~Courbin\orcid{0000-0003-0758-6510}\inst{\ref{aff59}}
\and H.~M.~Courtois\orcid{0000-0003-0509-1776}\inst{\ref{aff60}}
\and A.~Da~Silva\orcid{0000-0002-6385-1609}\inst{\ref{aff61},\ref{aff62}}
\and H.~Degaudenzi\orcid{0000-0002-5887-6799}\inst{\ref{aff13}}
\and A.~M.~Di~Giorgio\orcid{0000-0002-4767-2360}\inst{\ref{aff63}}
\and J.~Dinis\orcid{0000-0001-5075-1601}\inst{\ref{aff61},\ref{aff62}}
\and M.~Douspis\orcid{0000-0003-4203-3954}\inst{\ref{aff25}}
\and F.~Dubath\orcid{0000-0002-6533-2810}\inst{\ref{aff13}}
\and X.~Dupac\inst{\ref{aff28}}
\and A.~Ealet\orcid{0000-0003-3070-014X}\inst{\ref{aff58}}
\and M.~Farina\orcid{0000-0002-3089-7846}\inst{\ref{aff63}}
\and S.~Farrens\orcid{0000-0002-9594-9387}\inst{\ref{aff30}}
\and S.~Ferriol\inst{\ref{aff58}}
\and S.~Fotopoulou\orcid{0000-0002-9686-254X}\inst{\ref{aff64}}
\and M.~Frailis\orcid{0000-0002-7400-2135}\inst{\ref{aff65}}
\and E.~Franceschi\orcid{0000-0002-0585-6591}\inst{\ref{aff12}}
\and P.~Franzetti\inst{\ref{aff48}}
\and S.~Galeotta\orcid{0000-0002-3748-5115}\inst{\ref{aff65}}
\and W.~Gillard\orcid{0000-0003-4744-9748}\inst{\ref{aff66}}
\and B.~Gillis\orcid{0000-0002-4478-1270}\inst{\ref{aff23}}
\and C.~Giocoli\orcid{0000-0002-9590-7961}\inst{\ref{aff12},\ref{aff67}}
\and P.~G\'omez-Alvarez\orcid{0000-0002-8594-5358}\inst{\ref{aff68},\ref{aff28}}
\and A.~Grazian\orcid{0000-0002-5688-0663}\inst{\ref{aff20}}
\and F.~Grupp\inst{\ref{aff18},\ref{aff37}}
\and L.~Guzzo\orcid{0000-0001-8264-5192}\inst{\ref{aff69},\ref{aff34}}
\and S.~V.~H.~Haugan\orcid{0000-0001-9648-7260}\inst{\ref{aff70}}
\and J.~Hoar\inst{\ref{aff28}}
\and H.~Hoekstra\orcid{0000-0002-0641-3231}\inst{\ref{aff71}}
\and W.~Holmes\inst{\ref{aff9}}
\and I.~Hook\orcid{0000-0002-2960-978X}\inst{\ref{aff72}}
\and F.~Hormuth\inst{\ref{aff73}}
\and A.~Hornstrup\orcid{0000-0002-3363-0936}\inst{\ref{aff74},\ref{aff3}}
\and P.~Hudelot\inst{\ref{aff7}}
\and K.~Jahnke\orcid{0000-0003-3804-2137}\inst{\ref{aff11}}
\and M.~Jhabvala\inst{\ref{aff75}}
\and E.~Keih\"anen\orcid{0000-0003-1804-7715}\inst{\ref{aff76}}
\and S.~Kermiche\orcid{0000-0002-0302-5735}\inst{\ref{aff66}}
\and A.~Kiessling\orcid{0000-0002-2590-1273}\inst{\ref{aff9}}
\and T.~Kitching\orcid{0000-0002-4061-4598}\inst{\ref{aff77}}
\and B.~Kubik\orcid{0009-0006-5823-4880}\inst{\ref{aff58}}
\and M.~K\"ummel\orcid{0000-0003-2791-2117}\inst{\ref{aff37}}
\and M.~Kunz\orcid{0000-0002-3052-7394}\inst{\ref{aff78}}
\and H.~Kurki-Suonio\orcid{0000-0002-4618-3063}\inst{\ref{aff79},\ref{aff80}}
\and O.~Lahav\orcid{0000-0002-1134-9035}\inst{\ref{aff81}}
\and D.~Le~Mignant\orcid{0000-0002-5339-5515}\inst{\ref{aff22}}
\and S.~Ligori\orcid{0000-0003-4172-4606}\inst{\ref{aff38}}
\and P.~B.~Lilje\orcid{0000-0003-4324-7794}\inst{\ref{aff70}}
\and V.~Lindholm\orcid{0000-0003-2317-5471}\inst{\ref{aff79},\ref{aff80}}
\and I.~Lloro\inst{\ref{aff82}}
\and D.~Maino\inst{\ref{aff69},\ref{aff48},\ref{aff83}}
\and E.~Maiorano\orcid{0000-0003-2593-4355}\inst{\ref{aff12}}
\and O.~Mansutti\orcid{0000-0001-5758-4658}\inst{\ref{aff65}}
\and O.~Marggraf\orcid{0000-0001-7242-3852}\inst{\ref{aff84}}
\and K.~Markovic\orcid{0000-0001-6764-073X}\inst{\ref{aff9}}
\and N.~Martinet\orcid{0000-0003-2786-7790}\inst{\ref{aff22}}
\and F.~Marulli\orcid{0000-0002-8850-0303}\inst{\ref{aff85},\ref{aff12},\ref{aff36}}
\and R.~Massey\orcid{0000-0002-6085-3780}\inst{\ref{aff86}}
\and D.~C.~Masters\orcid{0000-0001-5382-6138}\inst{\ref{aff87}}
\and S.~Maurogordato\inst{\ref{aff88}}
\and H.~J.~McCracken\orcid{0000-0002-9489-7765}\inst{\ref{aff7}}
\and E.~Medinaceli\orcid{0000-0002-4040-7783}\inst{\ref{aff12}}
\and S.~Mei\orcid{0000-0002-2849-559X}\inst{\ref{aff89}}
\and M.~Melchior\inst{\ref{aff90}}
\and Y.~Mellier\inst{\ref{aff24},\ref{aff7}}
\and M.~Meneghetti\orcid{0000-0003-1225-7084}\inst{\ref{aff12},\ref{aff36}}
\and E.~Merlin\orcid{0000-0001-6870-8900}\inst{\ref{aff49}}
\and G.~Meylan\inst{\ref{aff59}}
\and J.~J.~Mohr\orcid{0000-0002-6875-2087}\inst{\ref{aff37},\ref{aff18}}
\and M.~Moresco\orcid{0000-0002-7616-7136}\inst{\ref{aff85},\ref{aff12}}
\and L.~Moscardini\orcid{0000-0002-3473-6716}\inst{\ref{aff85},\ref{aff12},\ref{aff36}}
\and R.~Nakajima\inst{\ref{aff84}}
\and R.~C.~Nichol\orcid{0000-0003-0939-6518}\inst{\ref{aff33}}
\and S.-M.~Niemi\inst{\ref{aff31}}
\and C.~Padilla\orcid{0000-0001-7951-0166}\inst{\ref{aff91}}
\and F.~Pasian\orcid{0000-0002-4869-3227}\inst{\ref{aff65}}
\and K.~Pedersen\inst{\ref{aff92}}
\and W.~J.~Percival\orcid{0000-0002-0644-5727}\inst{\ref{aff93},\ref{aff94},\ref{aff95}}
\and V.~Pettorino\inst{\ref{aff31}}
\and S.~Pires\orcid{0000-0002-0249-2104}\inst{\ref{aff30}}
\and G.~Polenta\orcid{0000-0003-4067-9196}\inst{\ref{aff96}}
\and M.~Poncet\inst{\ref{aff97}}
\and L.~A.~Popa\inst{\ref{aff98}}
\and L.~Pozzetti\orcid{0000-0001-7085-0412}\inst{\ref{aff12}}
\and F.~Raison\orcid{0000-0002-7819-6918}\inst{\ref{aff18}}
\and A.~Renzi\orcid{0000-0001-9856-1970}\inst{\ref{aff19},\ref{aff99}}
\and J.~Rhodes\orcid{0000-0002-4485-8549}\inst{\ref{aff9}}
\and G.~Riccio\inst{\ref{aff42}}
\and E.~Romelli\orcid{0000-0003-3069-9222}\inst{\ref{aff65}}
\and M.~Roncarelli\orcid{0000-0001-9587-7822}\inst{\ref{aff12}}
\and E.~Rossetti\orcid{0000-0003-0238-4047}\inst{\ref{aff35}}
\and R.~Saglia\orcid{0000-0003-0378-7032}\inst{\ref{aff37},\ref{aff18}}
\and D.~Sapone\orcid{0000-0001-7089-4503}\inst{\ref{aff100}}
\and P.~Schneider\orcid{0000-0001-8561-2679}\inst{\ref{aff84}}
\and T.~Schrabback\orcid{0000-0002-6987-7834}\inst{\ref{aff101}}
\and A.~Secroun\orcid{0000-0003-0505-3710}\inst{\ref{aff66}}
\and G.~Seidel\orcid{0000-0003-2907-353X}\inst{\ref{aff11}}
\and S.~Serrano\orcid{0000-0002-0211-2861}\inst{\ref{aff55},\ref{aff102},\ref{aff54}}
\and C.~Sirignano\orcid{0000-0002-0995-7146}\inst{\ref{aff19},\ref{aff99}}
\and G.~Sirri\orcid{0000-0003-2626-2853}\inst{\ref{aff36}}
\and L.~Stanco\orcid{0000-0002-9706-5104}\inst{\ref{aff99}}
\and P.~Tallada-Cresp\'{i}\orcid{0000-0002-1336-8328}\inst{\ref{aff51},\ref{aff52}}
\and A.~N.~Taylor\inst{\ref{aff23}}
\and H.~I.~Teplitz\orcid{0000-0002-7064-5424}\inst{\ref{aff87}}
\and I.~Tereno\inst{\ref{aff61},\ref{aff103}}
\and R.~Toledo-Moreo\orcid{0000-0002-2997-4859}\inst{\ref{aff104}}
\and I.~Tutusaus\orcid{0000-0002-3199-0399}\inst{\ref{aff105}}
\and L.~Valenziano\orcid{0000-0002-1170-0104}\inst{\ref{aff12},\ref{aff106}}
\and T.~Vassallo\orcid{0000-0001-6512-6358}\inst{\ref{aff37},\ref{aff65}}
\and A.~Veropalumbo\orcid{0000-0003-2387-1194}\inst{\ref{aff34},\ref{aff40},\ref{aff107}}
\and Y.~Wang\orcid{0000-0002-4749-2984}\inst{\ref{aff87}}
\and J.~Weller\orcid{0000-0002-8282-2010}\inst{\ref{aff37},\ref{aff18}}
\and E.~Zucca\orcid{0000-0002-5845-8132}\inst{\ref{aff12}}
\and C.~Burigana\orcid{0000-0002-3005-5796}\inst{\ref{aff108},\ref{aff106}}
\and G.~Castignani\orcid{0000-0001-6831-0687}\inst{\ref{aff12}}
\and Z.~Sakr\orcid{0000-0002-4823-3757}\inst{\ref{aff109},\ref{aff105},\ref{aff110}}
\and V.~Scottez\inst{\ref{aff24},\ref{aff111}}
\and M.~Viel\orcid{0000-0002-2642-5707}\inst{\ref{aff112},\ref{aff65},\ref{aff113},\ref{aff114},\ref{aff115}}
\and P.~Simon\inst{\ref{aff84}}
\and J.~Mart\'{i}n-Fleitas\orcid{0000-0002-8594-569X}\inst{\ref{aff116}}
\and D.~Scott\orcid{0000-0002-6878-9840}\inst{\ref{aff117}}}
										   
\institute{Department of Astronomy, University of Massachusetts, Amherst, MA 01003, USA\label{aff1}
\and
Physics and Astronomy Department, University of California, 900 University Ave., Riverside, CA 92521, USA\label{aff2}
\and
Cosmic Dawn Center (DAWN), Denmark\label{aff3}
\and
Niels Bohr Institute, University of Copenhagen, Jagtvej 128, 2200 Copenhagen, Denmark\label{aff4}
\and
Institute for Astronomy, University of Hawaii, 2680 Woodlawn Drive, Honolulu, HI 96822, USA\label{aff5}
\and
Cosmic Dawn Center (DAWN)\label{aff6}
\and
Institut d'Astrophysique de Paris, UMR 7095, CNRS, and Sorbonne Universit\'e, 98 bis boulevard Arago, 75014 Paris, France\label{aff7}
\and
Jodrell Bank Centre for Astrophysics, Department of Physics and Astronomy, University of Manchester, Oxford Road, Manchester M13 9PL, UK\label{aff8}
\and
Jet Propulsion Laboratory, California Institute of Technology, 4800 Oak Grove Drive, Pasadena, CA, 91109, USA\label{aff9}
\and
Department of Astronomy, School of Science, The University of Tokyo, 7-3-1 Hongo, Bunkyo, Tokyo 113-0033, Japan\label{aff10}
\and
Max-Planck-Institut f\"ur Astronomie, K\"onigstuhl 17, 69117 Heidelberg, Germany\label{aff11}
\and
INAF-Osservatorio di Astrofisica e Scienza dello Spazio di Bologna, Via Piero Gobetti 93/3, 40129 Bologna, Italy\label{aff12}
\and
Department of Astronomy, University of Geneva, ch. d'Ecogia 16, 1290 Versoix, Switzerland\label{aff13}
\and
School of Physical Sciences, The Open University, Milton Keynes, MK7 6AA, UK\label{aff14}
\and
Astrophysics Group, Blackett Laboratory, Imperial College London, London SW7 2AZ, UK\label{aff15}
\and
Department of Mathematics, Imperial College London, London SW7 2AZ, UK\label{aff16}
\and
School of Physics and Astronomy, University of Nottingham, University Park, Nottingham NG7 2RD, UK\label{aff17}
\and
Max Planck Institute for Extraterrestrial Physics, Giessenbachstr. 1, 85748 Garching, Germany\label{aff18}
\and
Dipartimento di Fisica e Astronomia "G. Galilei", Universit\`a di Padova, Via Marzolo 8, 35131 Padova, Italy\label{aff19}
\and
INAF-Osservatorio Astronomico di Padova, Via dell'Osservatorio 5, 35122 Padova, Italy\label{aff20}
\and
Instituto de F\'isica de Cantabria, Edificio Juan Jord\'a, Avenida de los Castros, 39005 Santander, Spain\label{aff21}
\and
Aix-Marseille Universit\'e, CNRS, CNES, LAM, Marseille, France\label{aff22}
\and
Institute for Astronomy, University of Edinburgh, Royal Observatory, Blackford Hill, Edinburgh EH9 3HJ, UK\label{aff23}
\and
Institut d'Astrophysique de Paris, 98bis Boulevard Arago, 75014, Paris, France\label{aff24}
\and
Universit\'e Paris-Saclay, CNRS, Institut d'astrophysique spatiale, 91405, Orsay, France\label{aff25}
\and
Observatoire Astronomique de Strasbourg (ObAS), Universit\'e de Strasbourg - CNRS, UMR 7550, Strasbourg, France\label{aff26}
\and
Kapteyn Astronomical Institute, University of Groningen, PO Box 800, 9700 AV Groningen, The Netherlands\label{aff27}
\and
ESAC/ESA, Camino Bajo del Castillo, s/n., Urb. Villafranca del Castillo, 28692 Villanueva de la Ca\~nada, Madrid, Spain\label{aff28}
\and
HE Space for European Space Agency (ESA), Camino bajo del Castillo, s/n, Urbanizacion Villafranca del Castillo, Villanueva de la Ca\~nada, 28692 Madrid, Spain\label{aff29}
\and
Universit\'e Paris-Saclay, Universit\'e Paris Cit\'e, CEA, CNRS, AIM, 91191, Gif-sur-Yvette, France\label{aff30}
\and
European Space Agency/ESTEC, Keplerlaan 1, 2201 AZ Noordwijk, The Netherlands\label{aff31}
\and
Department of Physics, Universit\'{e} de Montr\'{e}al, 2900 Edouard Montpetit Blvd, Montr\'{e}al, Qu\'{e}bec H3T 1J4, Canada\label{aff32}
\and
School of Mathematics and Physics, University of Surrey, Guildford, Surrey, GU2 7XH, UK\label{aff33}
\and
INAF-Osservatorio Astronomico di Brera, Via Brera 28, 20122 Milano, Italy\label{aff34}
\and
Dipartimento di Fisica e Astronomia, Universit\`a di Bologna, Via Gobetti 93/2, 40129 Bologna, Italy\label{aff35}
\and
INFN-Sezione di Bologna, Viale Berti Pichat 6/2, 40127 Bologna, Italy\label{aff36}
\and
Universit\"ats-Sternwarte M\"unchen, Fakult\"at f\"ur Physik, Ludwig-Maximilians-Universit\"at M\"unchen, Scheinerstrasse 1, 81679 M\"unchen, Germany\label{aff37}
\and
INAF-Osservatorio Astrofisico di Torino, Via Osservatorio 20, 10025 Pino Torinese (TO), Italy\label{aff38}
\and
Dipartimento di Fisica, Universit\`a di Genova, Via Dodecaneso 33, 16146, Genova, Italy\label{aff39}
\and
INFN-Sezione di Genova, Via Dodecaneso 33, 16146, Genova, Italy\label{aff40}
\and
Department of Physics "E. Pancini", University Federico II, Via Cinthia 6, 80126, Napoli, Italy\label{aff41}
\and
INAF-Osservatorio Astronomico di Capodimonte, Via Moiariello 16, 80131 Napoli, Italy\label{aff42}
\and
INFN section of Naples, Via Cinthia 6, 80126, Napoli, Italy\label{aff43}
\and
Instituto de Astrof\'isica e Ci\^encias do Espa\c{c}o, Universidade do Porto, CAUP, Rua das Estrelas, PT4150-762 Porto, Portugal\label{aff44}
\and
Faculdade de Ci\^encias da Universidade do Porto, Rua do Campo de Alegre, 4150-007 Porto, Portugal\label{aff45}
\and
Dipartimento di Fisica, Universit\`a degli Studi di Torino, Via P. Giuria 1, 10125 Torino, Italy\label{aff46}
\and
INFN-Sezione di Torino, Via P. Giuria 1, 10125 Torino, Italy\label{aff47}
\and
INAF-IASF Milano, Via Alfonso Corti 12, 20133 Milano, Italy\label{aff48}
\and
INAF-Osservatorio Astronomico di Roma, Via Frascati 33, 00078 Monteporzio Catone, Italy\label{aff49}
\and
INFN-Sezione di Roma, Piazzale Aldo Moro, 2 - c/o Dipartimento di Fisica, Edificio G. Marconi, 00185 Roma, Italy\label{aff50}
\and
Centro de Investigaciones Energ\'eticas, Medioambientales y Tecnol\'ogicas (CIEMAT), Avenida Complutense 40, 28040 Madrid, Spain\label{aff51}
\and
Port d'Informaci\'{o} Cient\'{i}fica, Campus UAB, C. Albareda s/n, 08193 Bellaterra (Barcelona), Spain\label{aff52}
\and
Institute for Theoretical Particle Physics and Cosmology (TTK), RWTH Aachen University, 52056 Aachen, Germany\label{aff53}
\and
Institute of Space Sciences (ICE, CSIC), Campus UAB, Carrer de Can Magrans, s/n, 08193 Barcelona, Spain\label{aff54}
\and
Institut d'Estudis Espacials de Catalunya (IEEC),  Edifici RDIT, Campus UPC, 08860 Castelldefels, Barcelona, Spain\label{aff55}
\and
Dipartimento di Fisica e Astronomia "Augusto Righi" - Alma Mater Studiorum Universit\`a di Bologna, Viale Berti Pichat 6/2, 40127 Bologna, Italy\label{aff56}
\and
European Space Agency/ESRIN, Largo Galileo Galilei 1, 00044 Frascati, Roma, Italy\label{aff57}
\and
Universit\'e Claude Bernard Lyon 1, CNRS/IN2P3, IP2I Lyon, UMR 5822, Villeurbanne, F-69100, France\label{aff58}
\and
Institute of Physics, Laboratory of Astrophysics, Ecole Polytechnique F\'ed\'erale de Lausanne (EPFL), Observatoire de Sauverny, 1290 Versoix, Switzerland\label{aff59}
\and
UCB Lyon 1, CNRS/IN2P3, IUF, IP2I Lyon, 4 rue Enrico Fermi, 69622 Villeurbanne, France\label{aff60}
\and
Departamento de F\'isica, Faculdade de Ci\^encias, Universidade de Lisboa, Edif\'icio C8, Campo Grande, PT1749-016 Lisboa, Portugal\label{aff61}
\and
Instituto de Astrof\'isica e Ci\^encias do Espa\c{c}o, Faculdade de Ci\^encias, Universidade de Lisboa, Campo Grande, 1749-016 Lisboa, Portugal\label{aff62}
\and
INAF-Istituto di Astrofisica e Planetologia Spaziali, via del Fosso del Cavaliere, 100, 00100 Roma, Italy\label{aff63}
\and
School of Physics, HH Wills Physics Laboratory, University of Bristol, Tyndall Avenue, Bristol, BS8 1TL, UK\label{aff64}
\and
INAF-Osservatorio Astronomico di Trieste, Via G. B. Tiepolo 11, 34143 Trieste, Italy\label{aff65}
\and
Aix-Marseille Universit\'e, CNRS/IN2P3, CPPM, Marseille, France\label{aff66}
\and
Istituto Nazionale di Fisica Nucleare, Sezione di Bologna, Via Irnerio 46, 40126 Bologna, Italy\label{aff67}
\and
FRACTAL S.L.N.E., calle Tulip\'an 2, Portal 13 1A, 28231, Las Rozas de Madrid, Spain\label{aff68}
\and
Dipartimento di Fisica "Aldo Pontremoli", Universit\`a degli Studi di Milano, Via Celoria 16, 20133 Milano, Italy\label{aff69}
\and
Institute of Theoretical Astrophysics, University of Oslo, P.O. Box 1029 Blindern, 0315 Oslo, Norway\label{aff70}
\and
Leiden Observatory, Leiden University, Einsteinweg 55, 2333 CC Leiden, The Netherlands\label{aff71}
\and
Department of Physics, Lancaster University, Lancaster, LA1 4YB, UK\label{aff72}
\and
Felix Hormuth Engineering, Goethestr. 17, 69181 Leimen, Germany\label{aff73}
\and
Technical University of Denmark, Elektrovej 327, 2800 Kgs. Lyngby, Denmark\label{aff74}
\and
NASA Goddard Space Flight Center, Greenbelt, MD 20771, USA\label{aff75}
\and
Department of Physics and Helsinki Institute of Physics, Gustaf H\"allstr\"omin katu 2, 00014 University of Helsinki, Finland\label{aff76}
\and
Mullard Space Science Laboratory, University College London, Holmbury St Mary, Dorking, Surrey RH5 6NT, UK\label{aff77}
\and
Universit\'e de Gen\`eve, D\'epartement de Physique Th\'eorique and Centre for Astroparticle Physics, 24 quai Ernest-Ansermet, CH-1211 Gen\`eve 4, Switzerland\label{aff78}
\and
Department of Physics, P.O. Box 64, 00014 University of Helsinki, Finland\label{aff79}
\and
Helsinki Institute of Physics, Gustaf H{\"a}llstr{\"o}min katu 2, University of Helsinki, Helsinki, Finland\label{aff80}
\and
Department of Physics and Astronomy, University College London, Gower Street, London WC1E 6BT, UK\label{aff81}
\and
NOVA optical infrared instrumentation group at ASTRON, Oude Hoogeveensedijk 4, 7991PD, Dwingeloo, The Netherlands\label{aff82}
\and
INFN-Sezione di Milano, Via Celoria 16, 20133 Milano, Italy\label{aff83}
\and
Universit\"at Bonn, Argelander-Institut f\"ur Astronomie, Auf dem H\"ugel 71, 53121 Bonn, Germany\label{aff84}
\and
Dipartimento di Fisica e Astronomia "Augusto Righi" - Alma Mater Studiorum Universit\`a di Bologna, via Piero Gobetti 93/2, 40129 Bologna, Italy\label{aff85}
\and
Department of Physics, Centre for Extragalactic Astronomy, Durham University, South Road, DH1 3LE, UK\label{aff86}
\and
Infrared Processing and Analysis Center, California Institute of Technology, Pasadena, CA 91125, USA\label{aff87}
\and
Universit\'e C\^{o}te d'Azur, Observatoire de la C\^{o}te d'Azur, CNRS, Laboratoire Lagrange, Bd de l'Observatoire, CS 34229, 06304 Nice cedex 4, France\label{aff88}
\and
Universit\'e Paris Cit\'e, CNRS, Astroparticule et Cosmologie, 75013 Paris, France\label{aff89}
\and
University of Applied Sciences and Arts of Northwestern Switzerland, School of Engineering, 5210 Windisch, Switzerland\label{aff90}
\and
Institut de F\'{i}sica d'Altes Energies (IFAE), The Barcelona Institute of Science and Technology, Campus UAB, 08193 Bellaterra (Barcelona), Spain\label{aff91}
\and
Department of Physics and Astronomy, University of Aarhus, Ny Munkegade 120, DK-8000 Aarhus C, Denmark\label{aff92}
\and
Waterloo Centre for Astrophysics, University of Waterloo, Waterloo, Ontario N2L 3G1, Canada\label{aff93}
\and
Department of Physics and Astronomy, University of Waterloo, Waterloo, Ontario N2L 3G1, Canada\label{aff94}
\and
Perimeter Institute for Theoretical Physics, Waterloo, Ontario N2L 2Y5, Canada\label{aff95}
\and
Space Science Data Center, Italian Space Agency, via del Politecnico snc, 00133 Roma, Italy\label{aff96}
\and
Centre National d'Etudes Spatiales -- Centre spatial de Toulouse, 18 avenue Edouard Belin, 31401 Toulouse Cedex 9, France\label{aff97}
\and
Institute of Space Science, Str. Atomistilor, nr. 409 M\u{a}gurele, Ilfov, 077125, Romania\label{aff98}
\and
INFN-Padova, Via Marzolo 8, 35131 Padova, Italy\label{aff99}
\and
Departamento de F\'isica, FCFM, Universidad de Chile, Blanco Encalada 2008, Santiago, Chile\label{aff100}
\and
Universit\"at Innsbruck, Institut f\"ur Astro- und Teilchenphysik, Technikerstr. 25/8, 6020 Innsbruck, Austria\label{aff101}
\and
Satlantis, University Science Park, Sede Bld 48940, Leioa-Bilbao, Spain\label{aff102}
\and
Instituto de Astrof\'isica e Ci\^encias do Espa\c{c}o, Faculdade de Ci\^encias, Universidade de Lisboa, Tapada da Ajuda, 1349-018 Lisboa, Portugal\label{aff103}
\and
Universidad Polit\'ecnica de Cartagena, Departamento de Electr\'onica y Tecnolog\'ia de Computadoras,  Plaza del Hospital 1, 30202 Cartagena, Spain\label{aff104}
\and
Institut de Recherche en Astrophysique et Plan\'etologie (IRAP), Universit\'e de Toulouse, CNRS, UPS, CNES, 14 Av. Edouard Belin, 31400 Toulouse, France\label{aff105}
\and
INFN-Bologna, Via Irnerio 46, 40126 Bologna, Italy\label{aff106}
\and
Dipartimento di Fisica, Universit\`a degli studi di Genova, and INFN-Sezione di Genova, via Dodecaneso 33, 16146, Genova, Italy\label{aff107}
\and
INAF, Istituto di Radioastronomia, Via Piero Gobetti 101, 40129 Bologna, Italy\label{aff108}
\and
Institut f\"ur Theoretische Physik, University of Heidelberg, Philosophenweg 16, 69120 Heidelberg, Germany\label{aff109}
\and
Universit\'e St Joseph; Faculty of Sciences, Beirut, Lebanon\label{aff110}
\and
Junia, EPA department, 41 Bd Vauban, 59800 Lille, France\label{aff111}
\and
IFPU, Institute for Fundamental Physics of the Universe, via Beirut 2, 34151 Trieste, Italy\label{aff112}
\and
SISSA, International School for Advanced Studies, Via Bonomea 265, 34136 Trieste TS, Italy\label{aff113}
\and
INFN, Sezione di Trieste, Via Valerio 2, 34127 Trieste TS, Italy\label{aff114}
\and
ICSC - Centro Nazionale di Ricerca in High Performance Computing, Big Data e Quantum Computing, Via Magnanelli 2, Bologna, Italy\label{aff115}
\and
Aurora Technology for European Space Agency (ESA), Camino bajo del Castillo, s/n, Urbanizacion Villafranca del Castillo, Villanueva de la Ca\~nada, 28692 Madrid, Spain\label{aff116}
\and
Department of Physics and Astronomy, University of British Columbia, Vancouver, BC V6T 1Z1, Canada\label{aff117}}    

%
%
\abstract{
This paper presents a search for high redshift galaxies from the \textit{Euclid} Early Release Observations program ``Magnifying Lens.'' The 1.5\,$\deg^2$ area covered by the twin Abell lensing cluster fields is comparable in size to the few other deep near-infrared surveys such as COSMOS, and so provides an opportunity to significantly increase known samples of rare UV-bright galaxies at $z\approx6$--8 ($M_{\rm UV}\lesssim-22$). Beyond their still uncertain role in reionisation, these UV-bright galaxies are ideal laboratories from which to study galaxy formation and constrain the bright-end of the UV luminosity function. Of the \num{501994} sources detected from a combined \YE, \JE, and \HE NISP detection image, 168 do not have any appreciable VIS/\IE\/ flux. These objects span a range in spectral colours, separated into two classes: 139 extremely red sources; and 29 Lyman-break galaxy candidates. Best-fit redshifts and spectral templates suggest the former is composed of both $z\gtrsim5$ dusty star-forming galaxies and $z\approx1$--3 quiescent systems. The latter is composed of more homogeneous Lyman break galaxies at $z\approx6$--8. In both cases, contamination by L- and T-type dwarfs cannot be ruled out with \Euclid images alone. Additional contamination from instrumental persistence is investigated using a novel time series analysis. This work lays the foundation for future searches within the Euclid Deep Fields, where thousands more $z\gtrsim6$ Lyman break systems and extremely red sources will be identified. 
}
%
%
    \keywords{Galaxies: high-redshift; Galaxies: evolution; Catalogues}
%
%
   \titlerunning{\Euclid: ERO -- NISP-only sources and the search for luminous $z=6$--8 galaxies}
   \authorrunning{Weaver et al.}
   
   \maketitle
%
%
%
%

\clearpage
   
\section{Introduction}\label{sc:Intro}

The reionisation of neutral hydrogen within the intergalactic medium (IGM) marks a major transition in cosmic history. Ionising Lyman continuum photons emitted from the first generations of massive stars is thought to have contributed significantly to the reionization budget \citep{Dayal2018,Finkelstein2019,Atek2024}. Observations from \textit{Planck} find $z=7.67\pm0.73$ as the mid-point of the reionization era, concluding sometime around $z\approx6$ with a transparent IGM \citep{PlanckCollaboration2016}. Within the standard paradigm of hierarchical structure formation, the first structures collapse within the most overdense regions of the primordial web of dark matter \citep{White1991, Behroozi2019}, from which the first stars form. As such, the topology of reionisation should be patchy, varying significantly over degree scales \citep{Trac2015, Neyer2023, Lu2024}. Galaxies born out of these overdensities are expected to be rapidly star-forming and therefore UV luminous, but correspondingly rare \citep{Kauffmann2022, Naidu2022}. Elsewhere, collections of less UV luminous systems formed in comparably greater numbers \citep{Bouwens2015,Finkelstein2015,Qin2019, Kauffmann2022, Bouwens2023, Leung2023, Adams2024, Donnan2024}. Recent advancements enabled by the \textit{James Webb} Space Telescope (JWST) suggest that while the total ionising flux of UV luminous galaxies is higher, the escape fraction of ionising photons and therefore relative ionising contribution may be subdominant \citep{Roberts-Borsani2023, Endsley2023, Atek2024}. Furthermore, the discovery of surprisingly massive active black holes at $z\approx6$--10 has re-ignited the possibility of a quasar contribution to reionisation \citep{Banados2018, Furtak2023b, juod2023, Kokorev2023, Dayal2024, Greene2024}. 

Owing to their apparent magnitudes $J\approx24$--26\,AB and observational accessibility from near-infrared (NIR) surveys, UV luminous galaxies ($M_{\rm UV}\lesssim -21$\,AB) from the reionization era ($z>6$) have been detected from the few degree-scale NIR surveys wide enough to find them, e.g., SXDF and COSMOS \citep{Bowler2014, Kauffmann2022, Donnan2023, Varadaraj2023}. 
Pre-JWST studies already hinted at a potential excess of UV-luminous galaxies at $z=8$--10 compared to typically assumed Schechter functional forms, and a seemingly slow evolution in the bright-end of the galaxy UV luminosity function at these redshifts \citep{Bowler2020}. This slow evolution in the number density of UV-bright galaxies is being confirmed by the most recent observations of JWST at $z>10$ \citep[e.g.][]{Harikane2023,Chemerynska2023,Adams2024}. However, their scarce number density on the sky makes it challenging to assemble large samples to accurately determine the shape of bright end of the UV luminosity function, which holds crucial information on the early assembly of galaxies. Neither the \textit{Hubble} Space Telescope (HST) nor JWST are capable of efficiently surveying degree-scale areas, requiring continued investment in large (albeit lower resolution) NIR surveys from space and the ground.

These UV luminous, rapidly star forming galaxies are expected to be some of the fastest growing systems of their epoch. Although speculative in nature, they are some of the most suitable candidates for being the progenitors of massive galaxies at later epochs, possibly also of the first generation to have ceased star-formation \citep{carnall23b, glazebrook24}. If so, then these rare systems are vital laboratories for examining galaxy evolution at its most extreme. Constraining their mass build up and future evolution requires star formation rates, stellar masses, star-formation histories, and estimates of their gas resevoirs. Detailed spectroscopic follow-up is only now beginning to reveal the extraordinary variation in their physical properties \citep{Endsley2022, Bowler2024, Algera2024, Schouws2023}. Given this diversity, the limited samples currently available from the few degree-scale deep NIR surveys are insufficient to characterise them as a population. Progress as to their nature and contribution to reionisation requires even larger NIR surveys.

Another major contribution from NIR surveys is the discovery of extremely red sources \citep[e.g.,][]{Wang2016, Franco2018, Wang2019, Zavala2021}. Presumably reddened by dust obscuration, far-infrared studies have accomplished much by identifying the UV light reprocessed by obscuring dust clouds. JWST too, in its first images, revealed a surprising abundance of extremely red, dust-obscured disc galaxies \citep{Nelson2023}, which may have significant contributions to the stellar mass and star-formation budgets at $z>3$ \citep{Gottumukkala2024, Wang2024, Williams2023}. However, quiescent galaxies with their characteristically old stellar populations are a known but important interloper population \citep{Barrufet2024}. Both pose significant obstacles to identifying genuine high-redshift galaxies \citep{Naidu2022b, Zavala2023}. 

The European Space Agency (ESA) \Euclid mission was launched in July 2023. A medium-class probe, it was designed to survey the Universe to uncover details as to the nature and evolution of dark matter and dark energy -- two elusive components of the $\Lambda$CDM cosmological model \citep{Laureijs11, EuclidSkyOverview}. Its two instruments are the visible instrument (VIS, \citealt{EuclidSkyVIS}) with its characteristic ultra-broad \IE passband ideally suited for obtaining high-resolution optical imaging necessary for weak lensing studies, and the Near-Infrared Spectrometer and Photometer (NISP, \citealt{EuclidSkyNISP}),  providing a complement of lower-resolution NIR data in the \YE, \JE, and \HE passbands \citep{Schirmer2022} to secure photometric redshifts in addition to slitless grism spectroscopy. This unique combination of instruments is key to achieving \Euclid's objectives, and goes hand-in-hand with the multi-tiered survey strategy ranging from the Euclid Wide Survey (EWS, \citealt{Scaramella-EP1}) over 14\,000\,deg$^{2}$ to a 53\,deg$^2$ Euclid Deep Survey (EDS) of three primary fields: the Euclid Deep Field North (EDF-North); the Euclid Deep Field South (EDF-South); and the Euclid Deep Field Fornax (EDF-Fornax). Additional complements from well-studied auxiliary fields, such as COSMOS and a self-calibration field within EDF-North, along with grism observations, are purposely designed to improve key measurements. 

Six Early Release Observations (ERO) projects were chosen to highlight the capability of \Euclid to study a variety of astrophysical phenomena \citep{EROData, EROcite}. Among them, ``Magnifying Lens'' (PI: H.~Atek) takes aim at two massive galaxy clusters, Abell~2390 and Abell~2764, with VIS/\IE, NISP/\YJHE,  as well as NISP grism observations \citep{EROLensData}. While the magnifying power of their deep gravitational potentials helps to resolve galaxies immediately behind the cluster, the large $0.75\,\deg^2$ area of each field enables an order-of-magnitude increase in the number of detectable $z\gtrsim6$ UV bright galaxies compared to small area blank fields from HST and JWST (e.g., CANDELS, \citealt{Duncan2014, Bouwens2015, Finkelstein2015}). This rich sample of high-redshift systems can then be used to constrain the UV luminosity functions of galaxies and quasars, the physics of which can be studied from detailed spectroscopic follow up. These same observations can also be used to constrain dark matter through weak lensing analyses, as well as the virial masses and assembly history of the cluster galaxies themselves.

This work aims to showcase the ability of \textit{Euclid} to identify rare NISP-only objects (i.e., not detected in VIS/\IE), such as luminous high-redshift galaxies and extremely red sources, and is organised as follows. Section~\ref{sec:cats} describes the observations, data reduction, and photometric catalogue of Abell~2390 and Abell~2764. Section~\ref{sec:sample} presents the methodology adopted to identify NISP-only objects. Section~\ref{sec:results} presents the photometrically-selected sample, its properties, sources of potential contamination, and physical interpretation. Section~\ref{sec:discussion} discusses the challenges encountered and a prospectus for identifying such systems from the EDS and auxiliary fields. These results are computed adopting a standard $\Lambda$CDM cosmology with $H_0=70$\,km\,s$^{-1}$\,Mpc$^{-1}$, $\Omega_{\rm m}=0.3$ and $\Omega_{\Lambda}=0.7$ throughout. All magnitudes are expressed in the AB system \citep{oke_absolute_1974}, for which a flux $f_\nu$ in $\mu$Jy ($10^{-29}$~erg~cm$^{-2}$s$^{-1}$Hz$^{-1}$) corresponds to an AB magnitude of $23.9-2.5\,\log_{10}(f_\nu/\mu{\rm Jy})$.

\section{Data and photometric catalogue}\label{sec:cats}

\subsection{Observations}
The ERO ``Magnifying Lens'' project consists of intermediate-depth observations of two known galaxy clusters, Abell~2390 and Abell~2764. While both are rich systems with many lensed background sources, Abell~2390 ($z=0.228$, \citealt{Sohn2020}) is better studied than Abell~2764 ($z=0.07$, \citealt{Katgert1996}). See \citet{EROLensData} for details of the field selection.

For each cluster, \textit{Euclid} obtained three reference observing sequences (ROS, see Fig.\,8 of \citealt{Scaramella-EP1}) of 70.2\,min which include VIS/\IE and NISP/\YJHE imaging, adopting a 3\arcmin$\times$3\arcmin~dither between each ROS to fill-in coverage between the 36 chips. Individual \IE, \YE, \JE, and \HE exposures were stacked to produce four deep mosaics using \texttt{SWarp} \citep{BertinSwarp} at native \ang{;;0.1}\,pix$^{-1}$ and \ang{;;0.3}\,pix$^{-1}$  scales, respectively. A 64\,pixel mesh and 3$\times$ smoothing factor was adopted to model and subtract the background light, which for VIS is more challenging given the bluer Galactic cirrus foreground. Although other images produced through other approaches were considered, this work adopts the ``Compact-sources'' images because they are optimised for photometry of faint, compact objects. See \citet{EROData} for a complete description of the reduction procedure.

Each mosaic spans $0.75\,\deg^2$, making the $\IE=27.1$\,AB (5$\sigma$ point source) VIS images some of the deepest at such high-resolution covering degree-scales, comparable only to the largest HST mosaic taken (ACS/F814W over 1.7\,deg$^{2}$ in COSMOS, \citealt{Koekemoer2011}). Similarly, the \YE, \JE, and $\HE\approx$24.5\,AB (5$\sigma$ point source) images are some of the deepest NIR imaging covering degree-scales, comparable only to VIDEO \citep{jarvis2013} and the UltraVISTA survey of COSMOS \citep{McCracken2012}. Although designed for identifying new strongly lensed galaxies, their large contiguous areas beyond the cluster centres make them well-suited for identifying high-redshift, rest-frame UV bright objects that are too rare to be found consistently by smaller NIR surveys.

\subsection{Photometric catalogue}
Sources were detected in each field using Pythonic Source Extractor (\texttt{SEP}, \citealt{Bertin1996, Barbary2016}) from a \texttt{CHI\_MEAN} coadd of all three NISP \YE, \JE, and \HE mosaics constructed using \texttt{SWarp} \citep{BertinSwarp}. To achieve maximal completeness, \texttt{SEP} was configured to detect sources whose pixels are significant above $1.5\,\sigma$ across at least three contiguous pixels after smoothing with a \ang{;;0.45} (1.5\,pixel) full-width-half-maximum Gaussian kernel (see \citealt{Zalesky2024} for relevant discussion). Aggressive deblending is crucial to separating the relatively low-resolution NISP-only sources from line-of-sight neighbours. As such, \texttt{SEP} is configured to deblend with $2\times10^5$ thresholds, requiring a contrast of $10^{-5}$, tuned by visual inspection. Cleaning was not applied as it is liable to remove real sources of interest. A total of \num{501994} sources were recovered (98\% have ${\rm S/N}>3$ in either \YE, \JE, or \HE) with approximately equal numbers in each 0.5\,deg$^{2}$ field: \num{252675} in Abell~2390; and \num{249319} in Abell~2764. 

Photometry was measured at source positions in circular apertures with diameters of 0\farcs1, 0\farcs2, 0\farcs3, 0\farcs6, 1\farcs0, 1\farcs2, 1\farcs5, and 2\arcsecf0. Uncertainties were measured similarly on the weight maps; their effective per-pixel error distributions were verified to be consistent with the per-pixel noise estimated directly from the science mosaics. Given the \ang{;;0.3}\,pix$^{-1}$ scale of NISP, aperture diameters below \ang{;;0.6} are generally only suitable for VIS photometry.

Considering the difference in resolution and spatial sampling between VIS and NISP, it would seem advantageous to convolve all of the images to the broadest PSF, which is \HE. However, this work exploits the unique potential of using VIS solely as a ``dropout'' band where the colour terms between \IE and \YE, \JE, and \HE are largely inconsequential relative to the much more important signal-to-noise ratio (S/N) estimate from VIS/\IE which would be severely degraded if the image were downsampled and smoothed. Instead, the VIS/\IE S/N can be optimally determined from the original high-resolution images using a suitably chosen aperture that is large enough to be robust to noise but not corrupted by neighbouring sources. This fact, together with the relatively similar PSFs between all three NISP bands and the expected compactness of high-redshift Lyman break galaxies (<1\arcsec in radius, e.g., \citealt{Bowler2017}) means that we can expect to produce high-fidelity NISP colours by simply assuming that any light outside the aperture of choice is explained by the PSF. 

All aperture flux estimates and their uncertainties are corrected to total flux by dividing by the fractional energy outside each aperture. This extended light distribution is measured from observed PSFs constructed by optimally stacking recentred cutouts of point sources identified from the science mosaics following \citet{Weaver2024}. Relevant multiplicative correction factors for VIS are 16$\times$, 3$\times$, and 2$\times$ for \ang{;;0.1}, \ang{;;0.2}, and \ang{;;0.3} diameters, respectively; and 1.2$\times$, 1.3$\times$, and for NISP are 1.4$\times$ for \ang{;;1.2} diameter apertures in \YE, \JE, and \HE, respectively. 

To facilitate community engagement with these ERO data, a combined photometric catalogue of all \num{501994} sources is released with this work, jointly with the overview paper of \citet{EROLensData}. The catalogue includes unique source identification numbers and field-specific identifiers, sky coordinates, elementary shape estimates, aperture and total fluxes, flux errors, and their aperture-to-total corrections, in addition to \texttt{Source~Extractor}-like flags. Photometry is provided already corrected for Galactic extinction mapped by \citet{Schlafly2011}, adopting the extinction curves of \citet{Fitzpatrick2007}. There is a large number of artefacts near the mosaic edges where only a single dither contributes, as well as along the diffraction spikes of bright stars; these sources are identified in a hand-built region mask that excludes 15\% of the total catalogue (\texttt{use\_phot}\,=\,0) near bright stars, diffraction spikes, and approximately 1\arcmin\,around the mosaic edge. For the complete details, consult the \texttt{README} document provided with the catalogue at \url{https://zenodo.org/doi/10.5281/zenodo.11151975}.

\section{Candidate identification}\label{sec:sample}

\begin{figure}
    \centering
    \includegraphics[width=1\linewidth]{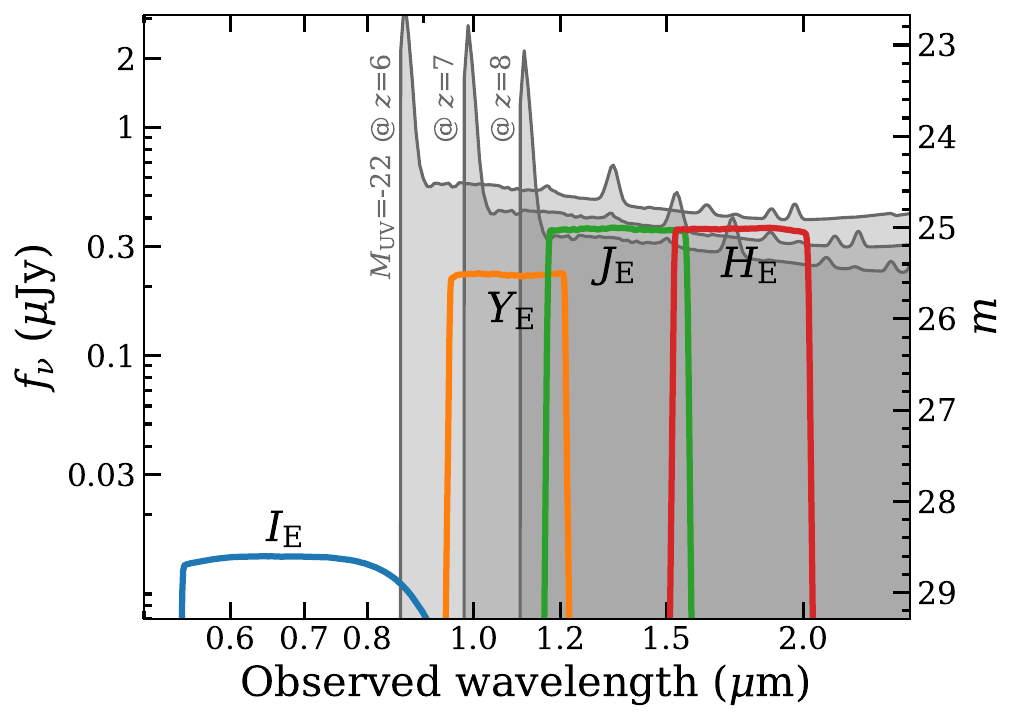}
    \caption{Summary of \Euclid's ability to identify bright $M_{\rm UV}<-22$ LBGs detectable only in NISP (or VIS/\IE ``dropout'') from the ERO data of Abell~2764 and Abell~2390. Model spectra of an LBG at $z=7$, 8, and 9 are generated with \texttt{python-FSPS}. \textit{Euclid} \IE, \YE, \JE, and \HE transmission curves are normalised such that their peak coincides with the selection function adopted in this work.
    }
    \label{fig:lbg_sed}
\end{figure}

Rather than only identifying galaxies at $z\approx6$--8, this work aims to identify all systems that are found only in the NISP \YE, \JE, and \HE bands, lacking any detectable VIS \IE signal. As such, a selection function is adopted requiring a ${\rm S/N}<1.5$ for VIS/\IE ($\approx28.5$ total) in each of 0\farcs1, 0\farcs2, and 0\farcs3 diameter apertures. While this unusually strict criterion can reject some real $z\approx6$--7 sources with non-negligible ionising UV continuum flux blueward of Ly$\alpha$, it is necessary in order to maximise reliability (future ancillary optical data will help to relax this). Furthermore, sources are required to have \JE and $\HE<25$ (${\rm S/N}\gtrsim5$) and $\YE<25.5$ (${\rm S/N}\gtrsim3$), all total magnitudes, relaxing the latter condition to allow for $z\approx8$ solutions where the Lyman break is only marginally observed, but limits the search to $z\lesssim8.5$. However, the criteria demand that even the fainest candidates have an extraordinarily strong break with $\IE-\YE>2$\,AB, greater than most Lyman-break selection criteria used in HST and JWST surveys \citep[e.g.,][]{Bouwens2015,Finkelstein2015,Atek2024, Adams2024}, making these candidates unusually secure. Figure~\ref{fig:lbg_sed} illustrates these criteria relative to an $M_{\rm UV}=-22$ LBG at $z=7$, 8, and 9 built from \texttt{Python-FSPS} \citep{Conroy2009,Conroy2010, Johnson2023}; this is a simple toy model assuming a maximally opaque foreground IGM, high Ly$\alpha$ escape, and no damping wing. The height of each transmission curve is set to the effective depth of this selection function.

For NISP, larger \ang{;;1.2} diameter apertures are chosen because they will enclose marginally resolved features expected of $z\approx6$--9 LBGs (e.g., \citealt{Bowler2017}). All colours are computed using the total fluxes because they produce less biased colours compared to aperture fluxes alone. To safeguard against artefacts near edges and diffraction spikes, sources in the masked regions are discounted. Furthermore, extremely extended sources unlikely to be high redshift are rejected by requiring that \HE fluxes in \ang{;;2.0} apertures contain less than 20\% more light than is contained in \ang{;;1.5} diameter apertures. Of the total 262 sources satisfying these criteria, a visual inspection identified 66 likely artefacts and 27 VIS detections arising from occasional unmasked anomalies in the mosaics, e.g., diffraction spikes, cosmic rays, and low exposure areas.

\begin{figure}
    \centering
    \includegraphics[width=1\linewidth]{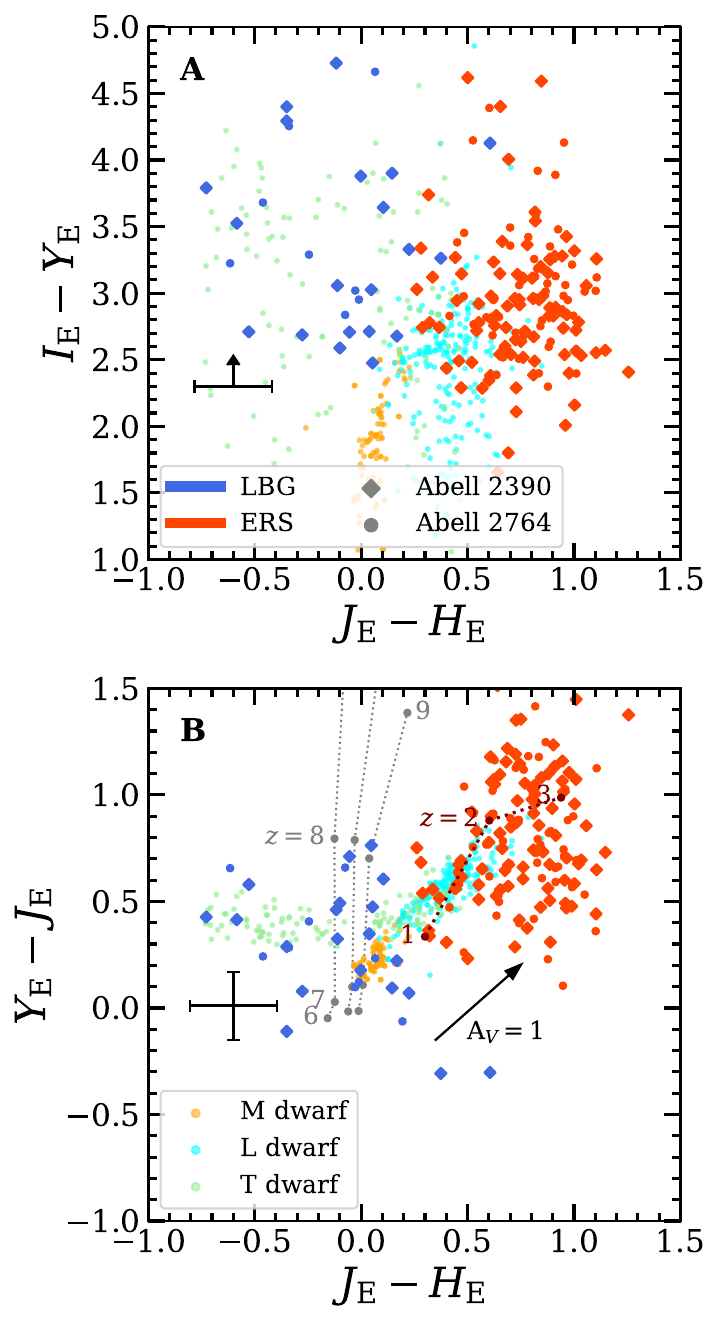}
    \caption{Selection and properties of 168 NISP-only sources in Abell~2390 and Abell~2764. Sources are divided into Lyman-break galaxy (LBG) candidates and extremely red sources (ERSs). \textit{Panel A}: $\IE-\YE$ versus $\JE-\HE$ colour-colour diagram highlighting the significance of the VIS non-detection ($1.5\sigma$ lower limits) and general colours of the sources. \textit{Panel B}: $\YE-\JE$ versus $\JE-\HE$ colour-colour diagram separating the two samples. Also shown is a reddening vector as well as tracks of $z=6$--9 LBGs with different mean stellar ages (grey) and of $z=1$--3 maximally old quiescent galaxies (maroon). Colours of observed M-, L-, and T-type dwarf stars are measured from the SpeX Prism Library \citep{Burgasser2004}. Typical colour uncertainties are shown in the upper two panels, where all $\IE-\YE$ colors are lower limits.
    }
    \label{fig:selection}
\end{figure}

\begin{figure}
    \centering
    \includegraphics[width=1\linewidth]{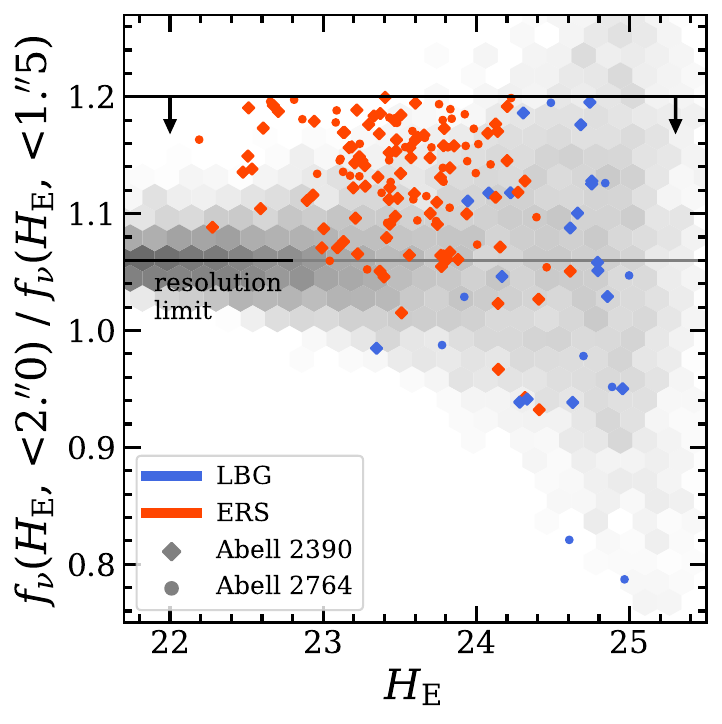}
    \caption{\HE brightness versus compactness estimated from a flux ratio in \ang{;;2.0} to \ang{;;1.5} apertures. Selected sources are required to be relatively compact with an aperture ratio below 1.2. The \HE compactness of genuine point sources identified in \IE from the parent catalogue is shown by the grey 2D histogram. Sources below a ratio of 1.0 are visually confirmed to have significantly negative pixels nearby.
    }
    \label{fig:compactness}
\end{figure}

Figure~\ref{fig:selection} and Fig.~\ref{fig:compactness} illustrate the colours and compactness of the 168 remaining candidates. Interestingly, despite overall similar detections in each field, 109 sources are found in Abell~2390 and only 59 in Abell~2764, a ratio of about 2 to 1. Reasons for this are likely related to brown dwarfs from the Galaxy, since the two fields have very different Galactic latitudes (see Sect.~\ref{sec:discussion}).

\section{Results}\label{sec:results}

The 168 sources selected in this work represent the most robust NISP-only objects chosen from Abell~2390 and Abell~2764. It is neither a complete sample nor a pure one, but is intended to showcase the potential of \textit{Euclid} and its ability to select promising high-redshift galaxies, among other interesting NISP-only sources.

\subsection{Properties of robust NISP-only sources}

The candidates span a remarkable range in colour from extremely red to relatively flat colours. Figure~\ref{fig:selection} provides an overview of the properties of the sample. To better contrast this diversity, we elect to refer to the redder subsample of 129 VIS-undetected objects as ``extremely red sources'' (ERSs), defined as having robustly red $\JE-H_{\rm E}$ and $\YE-\JE$ colours, whereby a flat $f_\nu$ spectrum cannot be supported within the 1$\sigma$ flux uncertainties of either colour. As evidenced by the reddening vector, these objects are consistent with high dust obscuration, although quiescent solutions are also possible (see Sect.~\ref{sec:discussion}). The remaining 29 objects with flat $f_\nu$ colours in both $\JE-H_{\rm E}$ and $\YE-\JE$ are then consistent with $z\approx6$--9 LBGs, as demonstrated by the grid of redshifts and mean stellar ages (10, 50, and 100\,Myr) derived from simple single stellar population models built with \texttt{Python-FSPS} including nebular emission line contributions. It is worth noting that although selected by consideration of their flux uncertainties, the two samples are clearly distinguished by their expected colours alone. Given that precise calibration of \textit{Euclid} is ongoing (although zero-point uncertainties are likely already below 0.1), this work purposely avoids defining a strict colour-colour selection for either category. Summary information of the LBG and ERS candidates are listed in Table~\ref{tab:app:lbg} and Table~\ref{tab:app:ers_1}--\ref{fig:app:ers_4}, respectively.

Morphologically, the majority of ERSs are at least marginally resolved or even disc-like whereas the LBGs occupy a range in compactness estimated from \HE flux ratios between \ang{;;1.5} and \ang{;;2.0} diameter apertures, as shown in Fig.~\ref{fig:selection}c. While the faintest LBGs appear more compact than point sources which have a typical flux ratio $\approx1.0$--1.1, they are likely unresolved objects whose flux ratio estimate is driven by noise. Visual inspection confirms this picture. Furthermore, only a few sources are close enough to the cluster core where magnification could be appreciable, which is particularly important for estimates of rest-frame UV magnitudes for the LBG candidates. Cutouts of all 168 sources, including false-colour images, are presented in Appendices~\ref{app:lbg_cutouts} and \ref{app:ero_cutouts}.

\subsection{Physical interpretations}

To learn more about the possible physical interpretation of these objects, best-fit spectral templates and photometric redshift likelihood distributions, $\mathcal{L}(z)$, were estimated by fitting the photometry using \texttt{eazy-py} \citep{Brammer2008}. To avoid contamination from neighbours, total \IE fluxes based on \ang{;;0.3} diameter apertures were coupled with total \YE, \JE, and \HE fluxes based on \ang{;;1.2} diameter apertures. At the time of writing, the most comprehensive template set provided with \texttt{eazy-py} is \texttt{sfhz\_blue\_agn}, which improves on previous template sets by including a physical prior on allowed star-formation histories to exclude quiescent templates for high-redshift solutions, adding an extreme emission line template suitable for $z>6$ UV-bright galaxies, and an active galactic nuclei torus template derived from \citet{Killi2023}. All priors were turned off, and iterative zero point corrections were not applied.

\begin{figure}
    \centering
    \includegraphics[width=1\linewidth]{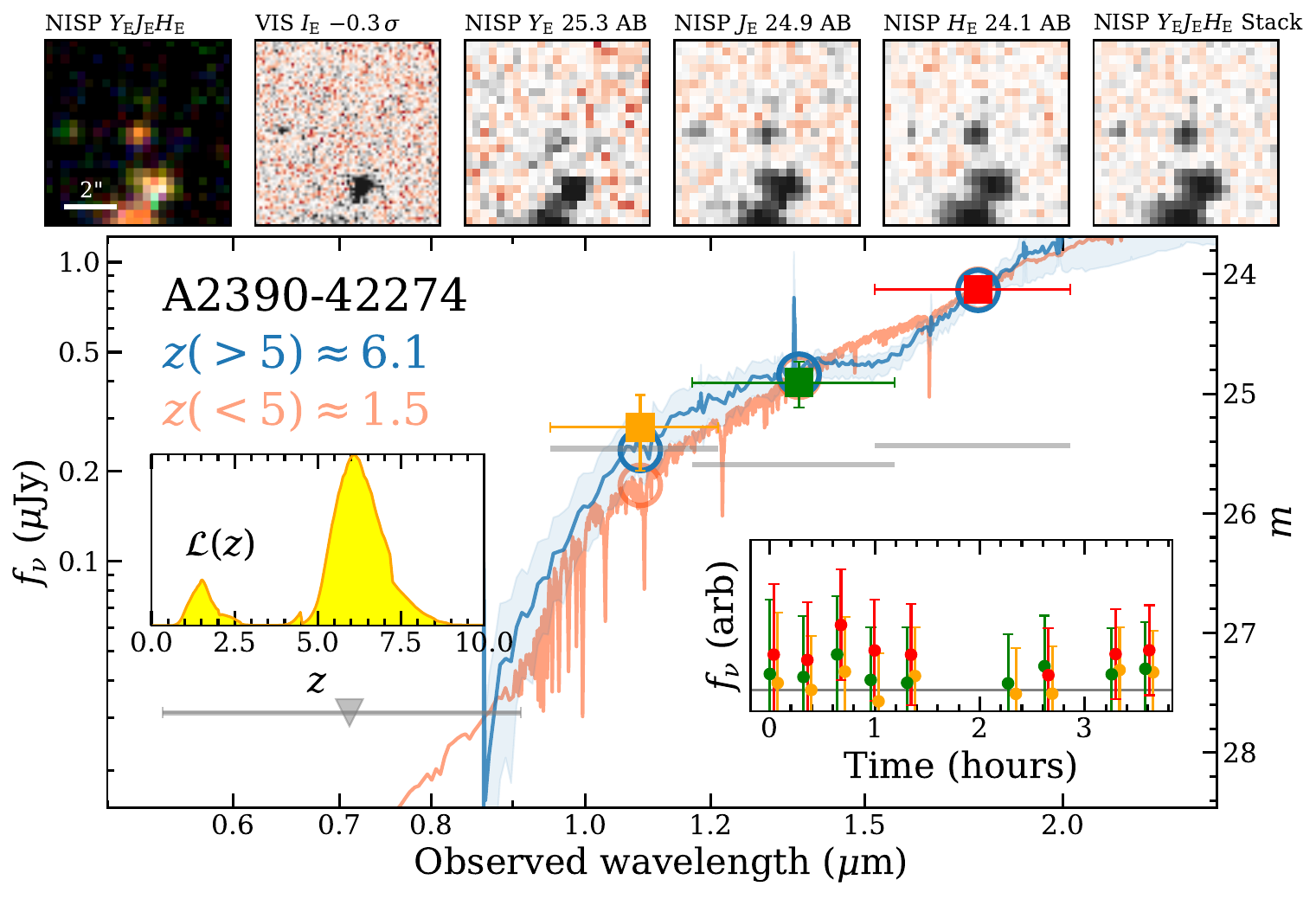}
    \includegraphics[width=1\linewidth]{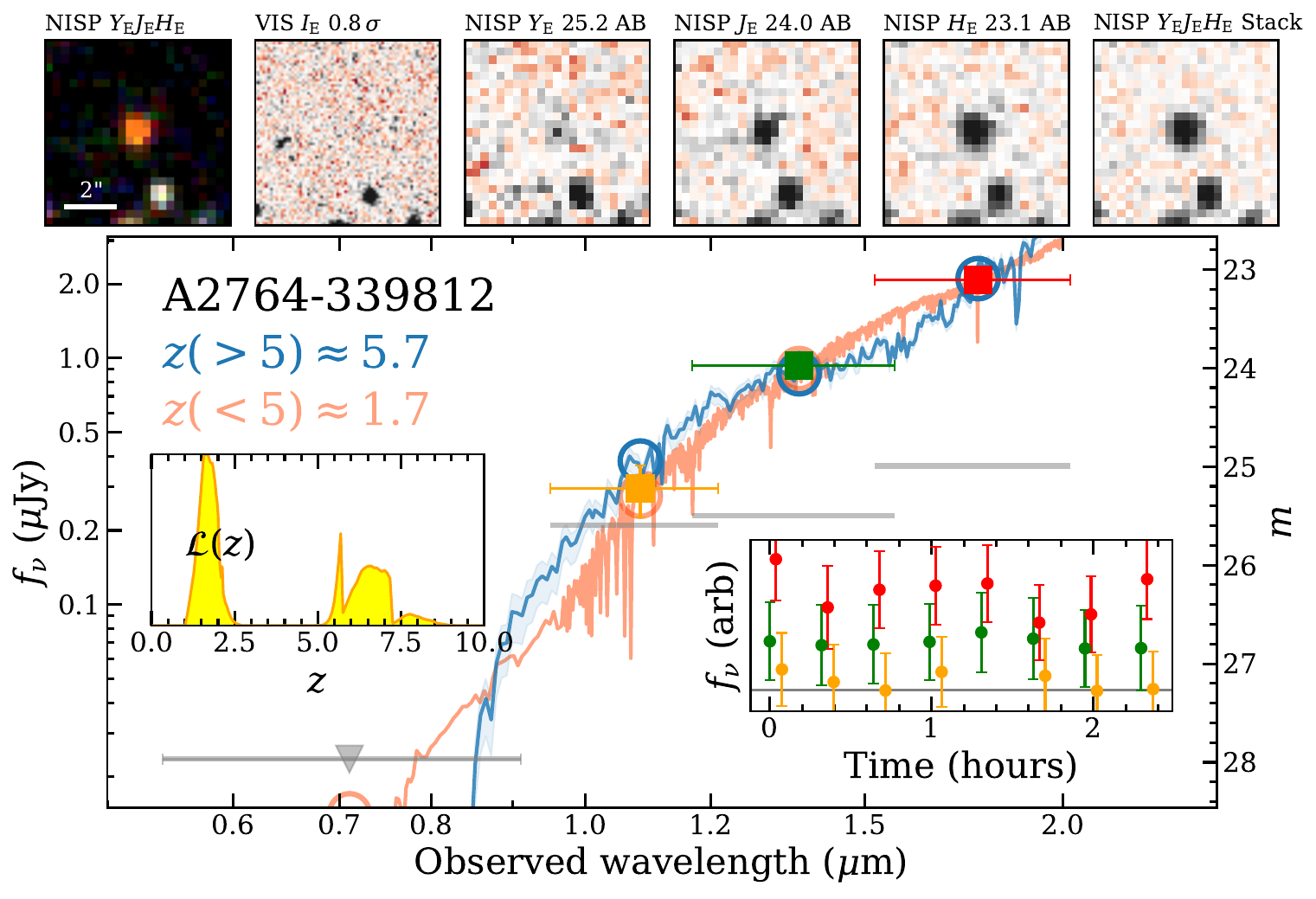}
    \caption{Two extremely red sources showing their photometry, best-fit templates and bimodal $\mathcal{L}(z)$ from \texttt{eazy-py}, cutouts, and light curves. The false colour image is composed from \YE, \JE, and \HE and the cutouts are scaled linearly such that they saturate at $\pm3\,\sigma$. Upper limits for \IE are shown by the leftmost grey bar with an arrow, while selection limits for NISP bands are shown as grey bars.
    }
    \label{fig:sed_ero}
\end{figure}

\begin{figure}
    \centering
    \includegraphics[width=1\linewidth]{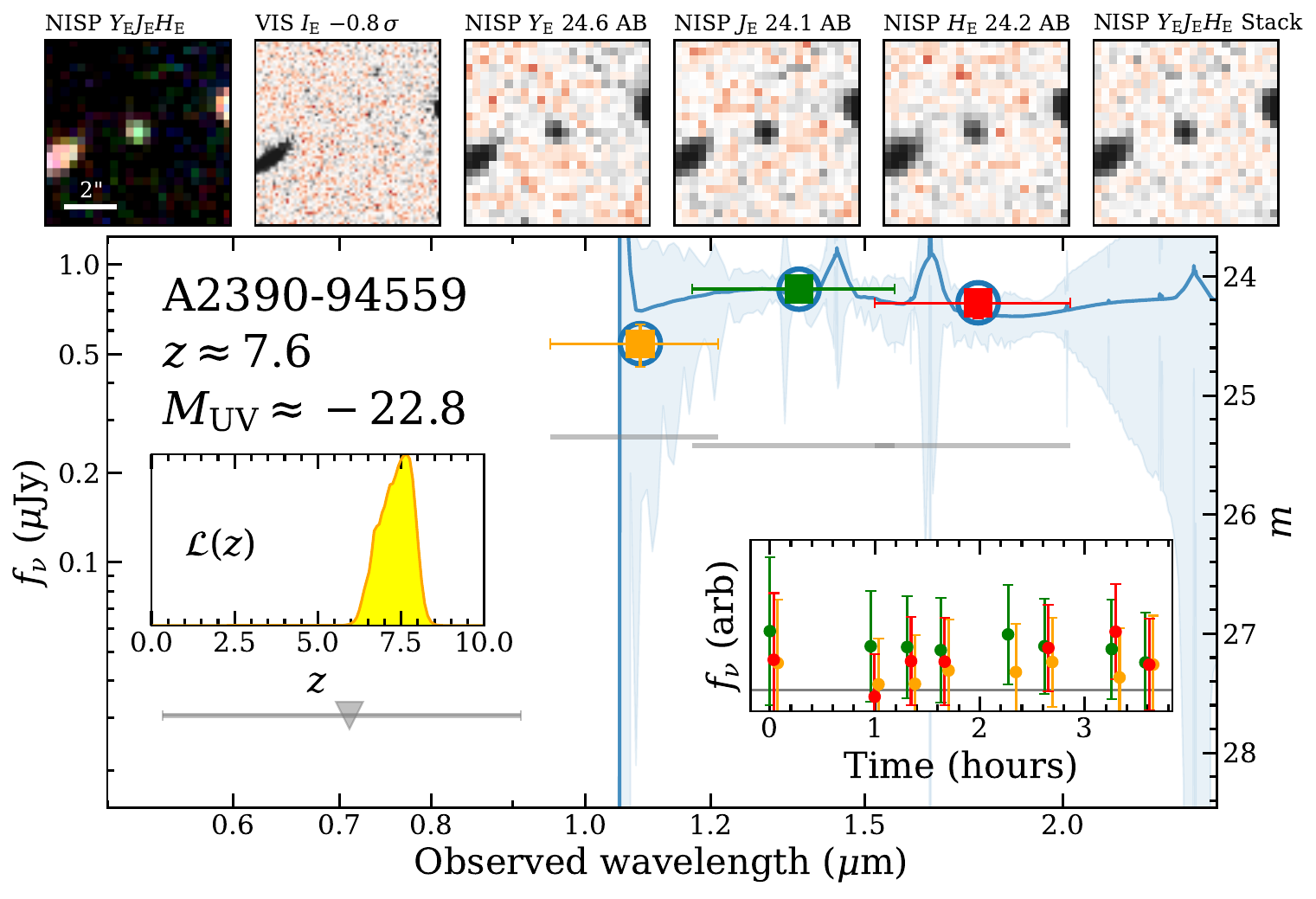}
    \includegraphics[width=1\linewidth]{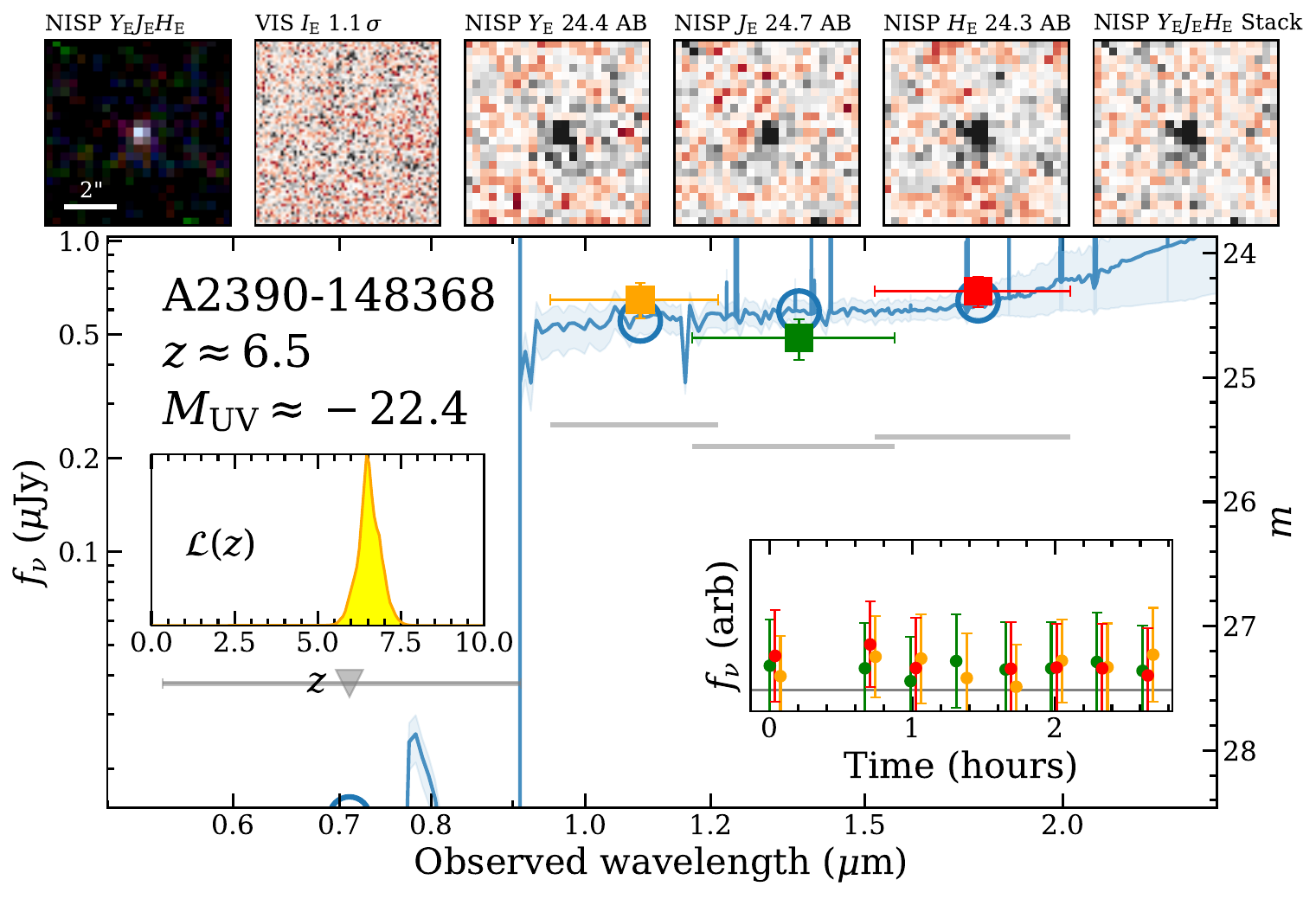}
    \caption{Two bright Lyman-break galaxy candidates: their photometry, best-fit templates and uniquely high-$z$ $\mathcal{L}(z)$ from \texttt{eazy-py}, cutouts, and lightcurves. The false colour image is composed of \YE, \JE, and \HE and the cutouts are scaled linearly such that they saturate at $\pm3\,\sigma$. Upper limits for \IE are shown by the leftmost grey bar with an arrow, while selection limits for NISP bands are shown by the grey bars.
    }
    \label{fig:sed_lbg}
\end{figure}

Having applied \texttt{eazy-py}, the differences between the two samples are clarified. While most of the ERSs have bimodal $\mathcal{L}(z)$ distributions, where a $z\approx1$--3 quiescent solution is degenerate with a $z\gtrsim5$ dusty solution, all but two LBG candidates have uniquely best-fit photometric redshift solutions at $z>6$ and best-fit templates consistent with their classification (the two exceptions are redder in $\JE-\HE$ and have some low-$z$ probability). Figure~\ref{fig:sed_ero} and \ref{fig:sed_lbg} show the best-fit template(s), $\mathcal{L}(z)$, and cutouts for two typical examples of ERSs and LBGs, respectively. Also shown are the frame-by-frame lightcurves extracted from each exposure (see Sect.\ref{sec:discussion} for details).

It is difficult to determine the nature of the ERSs conclusively, and it is probably a mixed sample. Contamination from brown dwarfs is likely for unresolved ERSs, given that L-type dwarfs exhibit similar \IE and \YE, \JE, and \HE colours (Fig.~\ref{fig:selection}). If extragalactic in nature, they could be maximally old quiescent systems at $z\approx1$--3 \citep{vanDokkum2008, Damjanov2009}, or extraordinarily dust-obscured star-forming galaxies at $z\gtrsim3$ with high stellar masses (e.g., \citealt{Manning2022, McKinney2023, Weaver2023_smf}). Figure~\ref{fig:sed_ero} highlights this uncertainty. In principle, deep, high-resolution far-infrared data could distinguish these possibilities, though this first-look study is solely focused on \textit{Euclid} and leaves further investigation of these sources for future work. This being said, a cursory inspection of ancillary \textit{Spitzer}/IRAC imaging of the center of Abell\,2390\footnote{Confirming these sources is difficult because HST coverage of Abell\,2390 is limited to the very core where no LBG or ERO source is found. Further, no ancillary HST or \textit{Spitzer} data exist for Abell\,2764.} detects every one of the six ERS targets they contain, confirming their extremely red nature.

The remaining LBG candidates, taken at face value, are extraordinarily bright objects for such an early epoch. Their rest-frame UV absolute magnitudes $M_{\rm UV}$ from \texttt{eazy-py} range from $-21.9$ to $-23.6$, comparable to similarly selected samples from UltraVISTA and VIDEO \citep{Bowler2017, Kauffmann2022, Varadaraj2023}. If genuine, these rapidly forming systems may be the progenitors of the most massive systems to evolve at later epochs of cosmic history. In addition, they should be valuable tracers of the overdensities in the cosmic web \citep{White1991}. Note that while the Abell clusters provide a magnification boost, most LBGs are found far from the cluster core and have magnification factors on the order of unity. Properly delensing the photometry and $M_{\rm UV}$ estimates requires a new lensing model that is not yet in hand.

Although an explicit calculation of the UV luminosity function is beyond the scope of this first-look study, the number densities of the 13  LBGs of $-23<M_{\rm UV}\leq-22$ at $6.5<z\leq7.5$ are comparable to literature estimates of the $z\sim7$ UV luminosity function of LBGs \citep[$\sim10$ over 1.5\,deg$^2$, e.g.,][]{Bowler2014, Finkelstein2015} or of quasars \citep{Schindler2023, Matsuoka2023}. This is encouraging given the uncertainties associated with photometry, redshifts, possible source magnification, and potential interlopers.

\section{Discussion}\label{sec:discussion}

Given that the EROs are only a first look at the \textit{Euclid} mission, the primary goal of this work is simply to present a sample of NISP-only sources, describe the challenges encountered, and discuss possible solutions to overcome them. As such, this work refrains from generalising this sample to population statistics (e.g. the UV luminosity function), leaving such investigation to near-future \textit{Euclid} studies where the 53\,deg$^2$ of Euclid Deep Survey and Auxillary Fields can be leveraged. 

\subsection{The expected appearance of high-$z$ sources}
\begin{figure}
    \centering
    \includegraphics[width=1\linewidth]{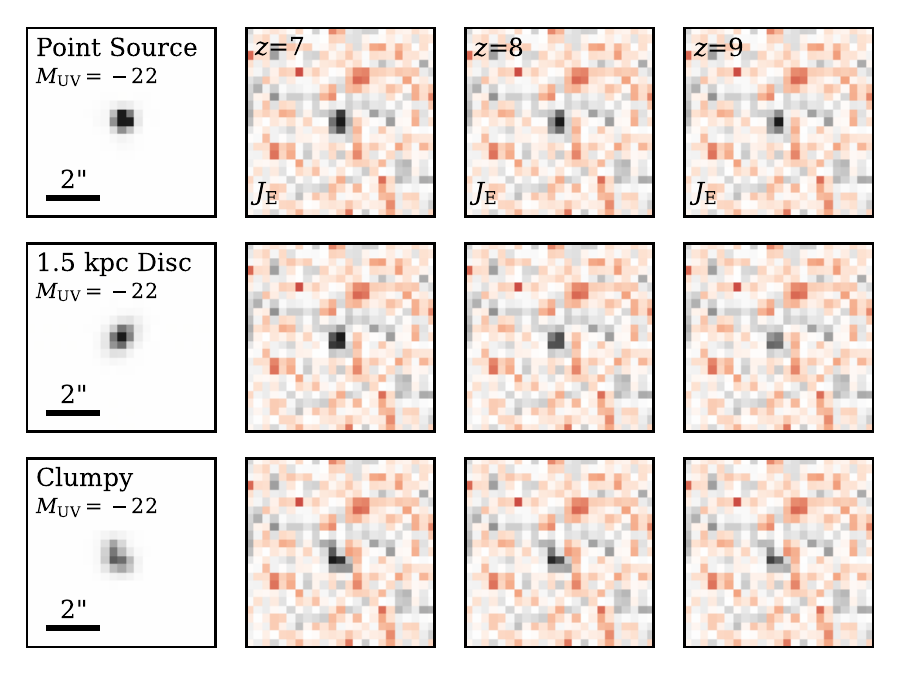}
    \caption{Models of a $M_{\rm UV}\approx-22$\,AB galaxy at $z=7$, 8, and 9 assuming three different morphologies: a point source; a 1.5\,kpc disc; and a three-point-source clump of similar effective size simulated with \texttt{The Tractor} \citep{Lang2016tractor} and injected into a small empty region of the \JE mosaic at native \ang{;;0.3}\,pix$^{-1}$ scale. 
    }
    \label{fig:simulation}
\end{figure}

Although VIS, with its finely sampled PSF, operates at the diffraction-limited resolution afforded by \textit{Euclid} outside the Earth's atmosphere, the broader PSF of NISP is still undersampled at a native \ang{;;0.3}\,pix$^{-1}$\, resolution, although ``drizzling'' can improve this. This effect is highly relevant for assessing morphologies of high-redshift galaxy candidates, since traditional techniques developed from e.g., HST, typically use compactness to separate real point-sources (e.g. stars) from marginally resolved compact galaxies. This technique cannot be applied to sources only found in NISP, where the superior resolution of VIS cannot be leveraged. Figure~\ref{fig:simulation} demonstrates this effect by injecting an $M_{\rm UV}=-22$ source into an empty region of the \HE image of Abell~2390 (with the same result in Abell~2764) with three light distributions: a point source; a 1.5\,kpc disc; and a tight clump of three point sources with a similar overall effective size of bright multi-component $z\approx7$ LBGs found by \citet{Bowler2017}. The observed morphology in NISP at \ang{;;0.3}\,pix$^{-1}$\, scale in the ERO data is virtually the same for all three cases, and by $z\approx9$ is indistinguishable given the low S/N. As such, no attempt was made to separate candidates from stars based on their observed morphologies, although this may be possible in the future by combining the frames on a higher-resolution grid (e.g. with ``drizzling'').

\subsection{The probability of brown dwarf contaminants}

\begin{figure}[b]
    \centering
    \includegraphics[width=0.9\linewidth]{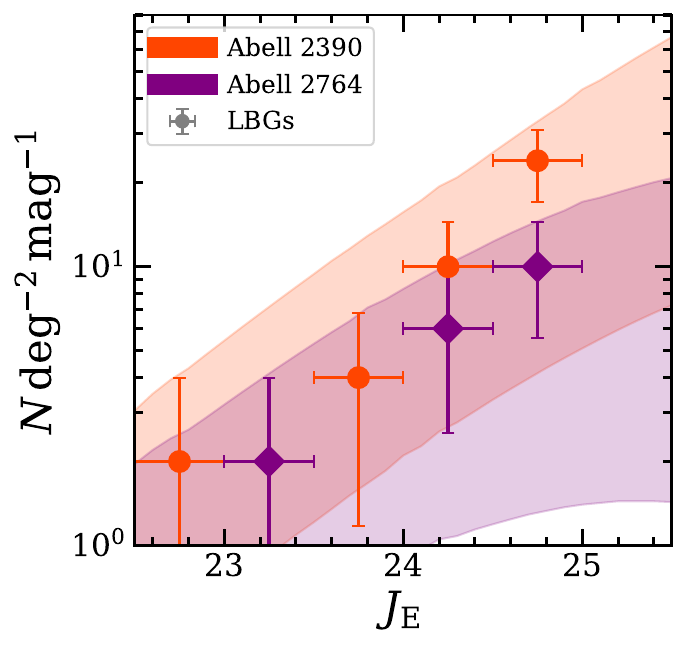}
    \caption{The number density of selected LBG candidates in Abell~2390 (orange diamonds) and Abell~2764 (purple diamons) compared to a range of T-type dwarf stars (700--1300\,K) number densities at the Galactic latitudes of each field. Dwarf sky densities are highly uncertain and are estimated assuming a Solar position of 27\,pc above the Galactic plane, an exponential radial scale height of 2250\,pc with a vertical scale length of 300\,pc \citep{Ryan2022}. See Wood et al., in prep. for details.
    }
    \label{fig:dwarfs}
\end{figure}

It is well known that searches for genuine high-redshift LBGs are complicated by late-type MLT dwarf stars because they have similar NIR colours and little optical flux \citep{Bowler2012, Bowler2014, Varadaraj2023, Harikane2022}. To illustrate this overlap, \IE, \YE, \JE, and \HE based colours were measured from observed spectra of M-, L-, and T-type dwarfs compiled in the SpeX prism spectral library of \citet{Burgasser2004} and are shown alongside the sample in Fig.~\ref{fig:selection} by orange, cyan, and green points, respectively. While the \YJHE colour-colour space shows that this sample is degenerate with M-, L-, and T-type dwarfs, the requirement of ${\rm S/N}<1.5$ in \IE and $\YE\gtrsim 25.5$ demands that candidates have an extraordinarily strong break with $\IE-\YE>2$. Furthermore, such a strong optical-NIR break is only expected from L- and T-type dwarfs, implying that the depth of \IE relative to that of \YE effectively excludes M-types by virtue of the ROS design. Furthermore, the factor of 2 discrepancy between the two fields is consistent with the expected (but highly uncertain) sky density of T-type dwarfs as illustrated in Fig.~\ref{fig:dwarfs}, suggesting that from a statistical standpoint that there may be a non-negligible population of T-type dwarf interlopers (see Wood et al., in prep.). In absence of suitably deep ancillary data, a definitive characterisation of these ERO sources requires follow-up spectroscopy and will have to wait for future work.

\subsection{Identifying cases of persistence}

The \textit{Euclid} mission follows a consistent survey pattern wherein a single visit consists of four dithered pointings with simultaneous observing by VIS and NISP enabled by splitting the primary beam with a dichroic filter (see \citealt{Scaramella-EP1} for details). While VIS integrates for an ultra-deep image ($\IE\approx 26$ per ROS), NISP begins with a 574-s integration of one of the two grism filters followed in order by equal 112-s integrations of \JE, \HE, and then \YE ($\YJHE\approx25.4$ per ROS). The order is important, since the NIR detectors of NISP are susceptible to charge persistence. This is where some charge from a previous exposure is trapped within the lattice defects of the NISP pixels. This manifests as a faint imprint of the previous exposures (grism or photometry) taken as much as 6-hours earlier. 

The persistence signal within the EROs was modelled and subtracted from each ERO exposure.
The persistence model was based on the flux of unsaturated pixels within all exposures taken up to 1\,hour before the ERO exposure; see Sect.~4.3.1 of \citet{EROData} for details. However, strong persistence (>70\,ADU) was not modelled because it would leave visible residuals. Therefore, we checked whether this left-over persistence impacted our detection of ERSs and LBGs.

\begin{figure}
    \centering
    \includegraphics[width=1\linewidth]{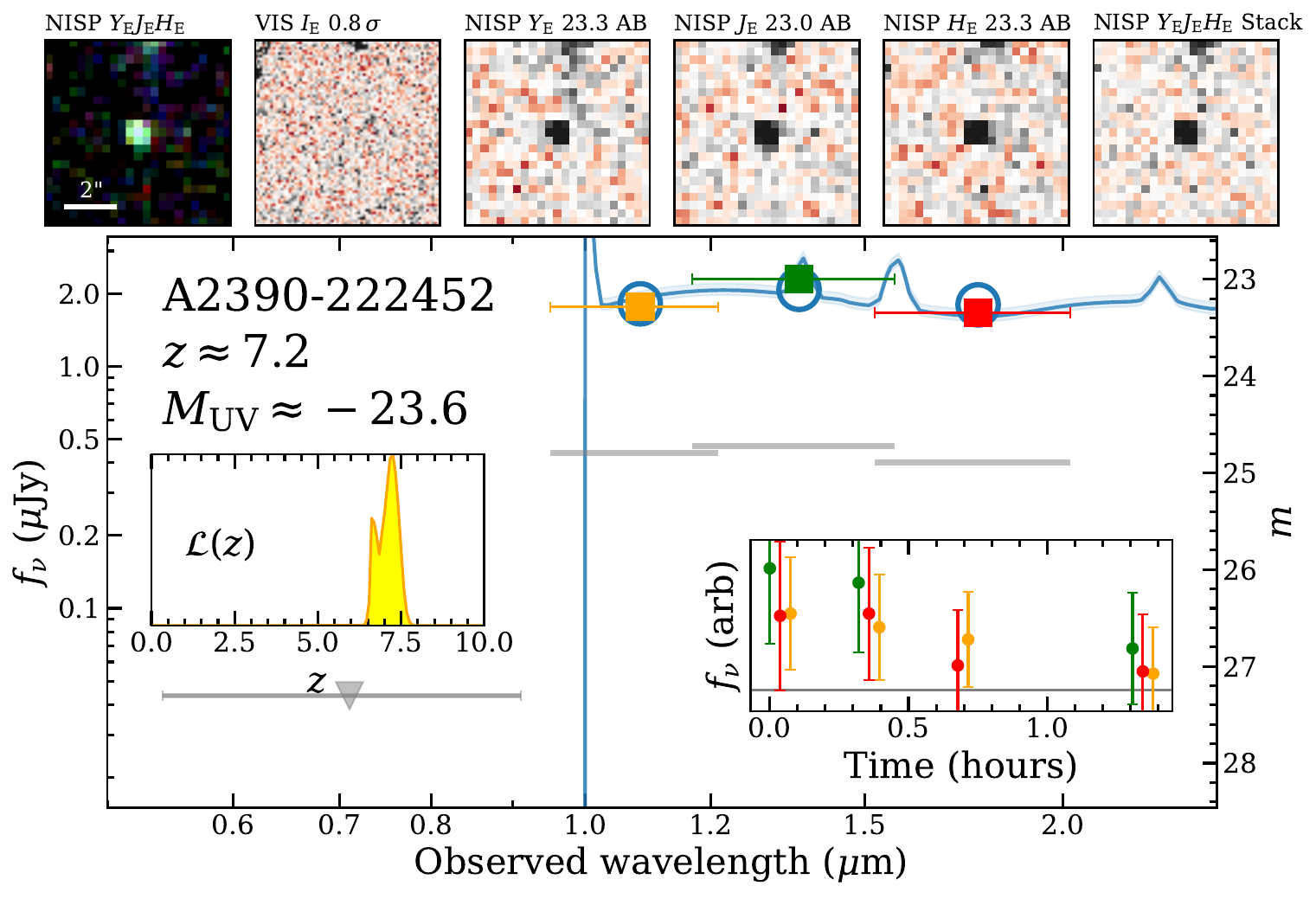}
    \caption{Example of a bright quasar-like object seen only in \YE, \JE, and \HE. Time-series analysis of its frame-by-frame photometry reveals a decaying signal suggestive of persistence.
    }
    \label{fig:persistence}
\end{figure}

Most of the brightest identifiable spectral persistence was successfully identified, modelled, and subtracted following the procedure developed specifically for the handling of the EROs. Furthermore, since the morphology of the persistence introduced by a stimulus from grism observations follows that of the original spectral trace of the object, the compactness criteria outlined in Sect.~\ref{sec:sample} effectively rejects the largest of remaining artefacts. However, there are some possible situations in which persistence may remain in the exposures and impact the detection of ERSs and LBGs. One such situation is persistence from cosmic rays in the same ROS sequence (which originally saturated the detector). Another persistence source is saturated compact sources from observations of a previous ROS. Since the persistence signal, $P$, is expected to decay approximately like $P_{\JE}\approx f_\nu/200$, $P_{\HE}\approx P_{\JE}/2$, and $P_{\YE}\approx P_{\JE}/3$ for a typical exposure, the spectral colours $\JE<\HE<\YE$ expected of persistence can mimic a $z\gtrsim6$ galaxy with appreciable Ly$\alpha$ emission or an extremely steep UV slope.

To facilitate the identification of such artefacts, light curves are constructed for each source by measuring their photometry in each pointing from all 12 exposures. Cosmic ray rejection and persistence cleaning are applied self-consistently to the mosaics and per-detector backgrounds are subtracted as $3\sigma$-clipped medians. Lacking calibrated weight maps, on-the-fly photometric uncertainties are estimated by the standard deviation of the fluxes from 10\,000 empty apertures placed around each image. 

An example of suspected persistence is shown in Fig.~\ref{fig:persistence}. If genuine, the stimulus occurred before or at the start of the observations of Abell~2390, unlike the spectro-to-photo persistence case mentioned above. However, while lightcurves are useful to rule out persistence in the case of bright sources, faint objects that are undetected in individual exposures ($\YJHE\approx 24$) will remain unidentifiable from a time series analysis. Given the expected exponential form of the UV luminosity function, the most common class of LBGs will be sufficiently faint to avoid detection in individual frames. While there were more obvious cases of persistence in earlier reductions that necessitated this lightcurve analysis, the removal of persistence in the ERO data improved significantly and the number of suspected cases diminished. In the final reductions we found no more extreme case than the candidate showed in Fig~\ref{fig:persistence}, and so without absolute certainly declined to remove other candidates. Thankfully, the existence of complementary data from multiple optical and infrared facilities will greatly improve the prospects for identifying NISP-only artefacts (see the section below for details).

\subsection{Outlook for Euclid Deep Fields}

Upon their completion, the Euclid Deep Fields (EDFs) will survey 53\,deg$^2$ across three fields; EDF-North, EDF-South, and EDF-Fornax, at 2 magnitudes deeper than the EWS \citep{Scaramella-EP1}, with these two Abell ERO fields of intermediate depth. The 53-fold increase in area and greater depth translates to more than 10\,000 LBGs at $z>7$ and up to 1000 at $z>8.5$ (see \citealt{Bowler2017}), providing the first statistically meaningful samples of rare, UV-luminous galaxies in the reionisation era. Such samples will finally enable constraints to be placed on the evolution of the bright-end of the UV luminosity function, among other investigations. The lessons learned from this work can be applied to the EDFs, since the visit pattern, pointings, and filter order are the same. The only difference is the greater number of visits (i.e., depth), along with more dither positions and position angles which, will improve persistence identification. The methodology laid out in this work is only the start, and will be refined as more data are taken over the survey lifetime.

Two fields in particular, EDF-North and EDF-Fornax, enjoy significant ancillary observations in the UV, optical, and infrared with contributions from both the Hawaii Two-0 Survey and Spitzer Legacy Survey \citep{Moneti-EP17}, forming, in conjunction with \Euclid, a part of the Cosmic Dawn Survey (DAWN). The availability of deep optical data from CFHT Megacam/$U$ and Subaru HSC/$grizy$ will improve the robustness of the candidates found, reveal more about their physical properties, and extend drop-out searches to lower redshifts. Observed over 6000 hours, the Spitzer Legacy Survey provides deep 3.6- and 4.5-$\mu$m IRAC imaging necessary to constrain the young stellar populations of $z\gtrsim3$ galaxies, when \HE becomes a rest-UV indicator. Furthermore, detection of NISP-only sources by IRAC rules out artefacts such as persistence and gives a lever-arm to better identify contaminating brown dwarfs, thereby greatly improving the purity of future LBG and ERS samples \citep{vanMierlo-EP21}. 

Zalesky et al. (in prep.) has established pre-\textit{Euclid} optically-selected photometric redshift catalogues in EDF-North and EDF-Fornax. While apertures used in the present work are highly effective, obtaining robust photometry of IRAC sources at their characteristically low resolution benefits significantly from fitting galaxy light profiles, e.g. with \texttt{The Farmer} \citep{Weaver2023_farmer}. With the addition of \IE, \YE, \JE, and \HE data, DAWN will provide NIR-selected catalogues of some 20\,million sources across EDF-North and EDF-Fornax with robust photometric redshifts and galaxy stellar masses out to $z\approx10$. Additional complementary data from ongoing spectroscopic campaigns from Keck/DEIMOS and Keck/MOSFIRE, together with the invaluable NISP grism observations, will provide excellent calibration of photometric redshifts and derived physical properties. The unprecedented statistical power achieved by combining \textit{Euclid}, CFHT, Subaru, and \textit{Spitzer} will establish EDF-North and EDF-Fornax as the leading extragalactic fields visible from the northern hemisphere for the next decade. Rich volume-complete samples of LBGs at $z=6$--10 will provide definitive constraints on the number density and evolution of UV-luminous galaxies necessary to firmly challenge and ultimately refine theories of galaxy formation.

\section{Summary}\label{sec:summary}

This paper presents one of the first science results from \Euclid, leveraging the deep degree-scale NIR imaging with NISP to identify rare, UV-bright LBGs and extremely red sources that are not seen in smaller blank fields. Our larger aim to obtain large samples to provide the statistical power from which their abundance, redshifts, and physical properties can be definitively studied.

Over half a million sources are detected from the twin $0.75\,\deg^2$ \YE, \JE, and \HE images, and their \IE, \YE, \JE, and \HE photometry is extracted from apertures corrected to total flux. Photometric redshifts and best-fit galaxy templates are estimated from \texttt{eazy-py}. The result is a highly bimodal sample spanning a range of spectral colours, with the following two classes of object.

\begin{itemize}
    \item 29 sources have flat $f_\nu$ colours indicative of $z\approx6$--8 Lyman break galaxies. Without magnification estimates, their $M_{\rm UV}$ spans $-21.9$ to $-23.6$, making this one of the largest samples of UV-luminous high-redshift galaxy candidates found thus far. Contamination by quasars and T-type dwarfs is likely and still needs to be determined.
    \item 139 sources have extremely red colors indicative of either $z\gtrsim5$ dusty star-forming galaxies or $z\approx1$--3 quiescent galaxies. Their redshift distributions are multi-modal, suggesting a mixture. A minority are point-like, which may be contamination by L-type dwarfs.
\end{itemize}

By selection, these objects are relatively compact, as anticipated for high-redshift LBGs. However, the morphological complexity of real LBGs seen with HST is less distinguishable here due to the \ang{;;0.3}\,pix$^{-1}$ scale of NISP. As a result, artefacts, such as instrumental persistence, cannot be identified from their morphologies. Instead, this work explores the novel use of time series analysis to identify characteristics unique to persistence from the lightcurves of suspected sources. This will be applicable to future \Euclid observations.

These comparably large samples of NISP-only sources will be dwarfed by those found imminently from the Euclid Deep Fields. EDF-North and EDF-Fornax, in particular, have deep optical Subaru/HSC and \textit{Spitzer}/IRAC coverage, which not only provides a longer lever-arm to identify brown dwarfs, but also a means to immediately identify artefacts arising only in NISP. Upon its completion, the Subaru-\Euclid-\textit{Spitzer} Cosmic Dawn Survey will identify thousands of similarly UV-luminous galaxies at $z>6$. These and other future searches will build on lessons learned from this work -- the first foray into the distant Universe with \Euclid.

%
%

\begin{acknowledgements}
The Cosmic Dawn Center (DAWN) is funded by the Danish National Research Foundation (DNRF140). HA is supported by the French Centre National d'Etudes Spatial (CNES). This work has made use of the \texttt{CANDIDE} Cluster at the \textit{Institut d'Astrophysique de Paris} (IAP), made possible by grants from the PNCG and the region of Île de France through the program DIM-ACAV+, and the Cosmic Dawn Center and maintained by S. Rouberol. CS acknowledges the support of the NSERC Postdoctoral Fellowship and the CITA National Fellowship programs.
This research has benefited from the SpeX Prism Spectral Libraries, maintained by Adam Burgasser at \url{http://www.browndwarfs.org/spexprism}.
This work made use of Astropy (\url{http://www.astropy.org}): a community-developed core Python package and an ecosystem of tools and resources for astronomy \citep{astropy:2013, astropy:2018, astropy:2022} and Matplotlib \citep{matplotlib07}.
\AckERO
\AckEC

\end{acknowledgements}

%
%

\bibliography{references, Euclid, EROplus}

%

\begin{appendix}
  \onecolumn 
  
\section{Lyman-break galaxy candidates}
\label{app:lbg_cutouts}

\begin{figure}%
    \centering
    \begin{subfigure}[T]{0.5\textwidth}
        \centering
        \includegraphics[width=0.92\textwidth]{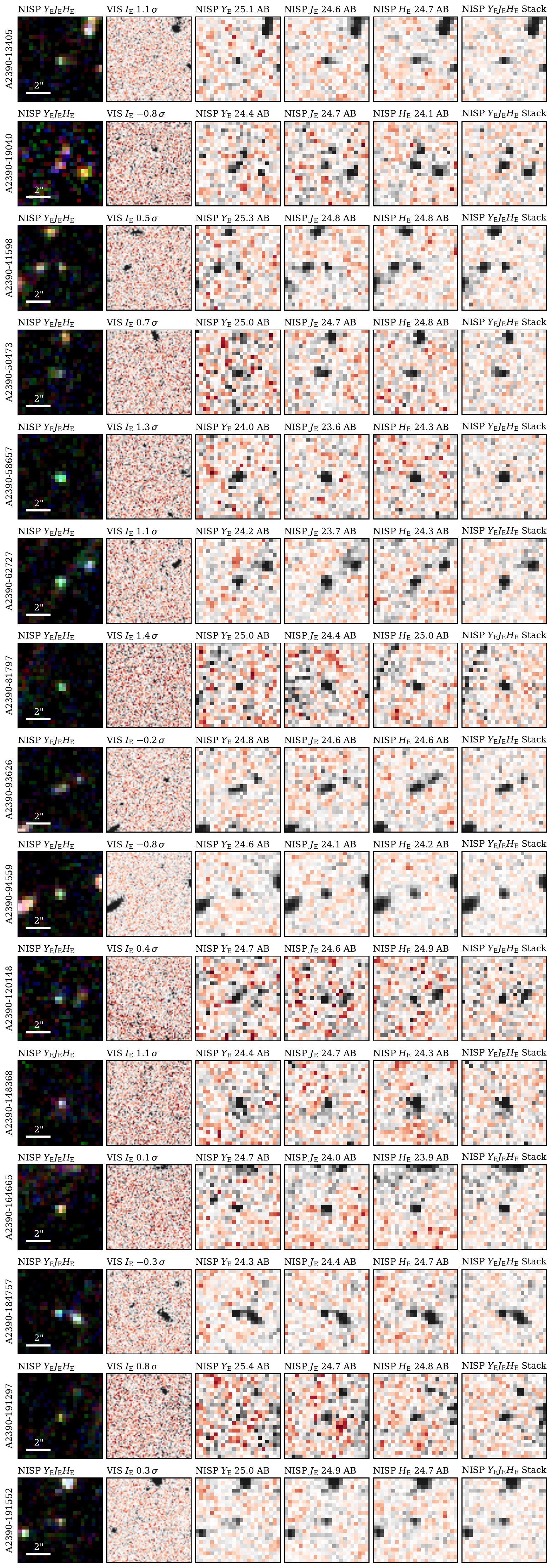}
    \end{subfigure}%
    ~ 
    \begin{subfigure}[T]{0.5\textwidth}
        \centering
        \includegraphics[width=0.92\textwidth]{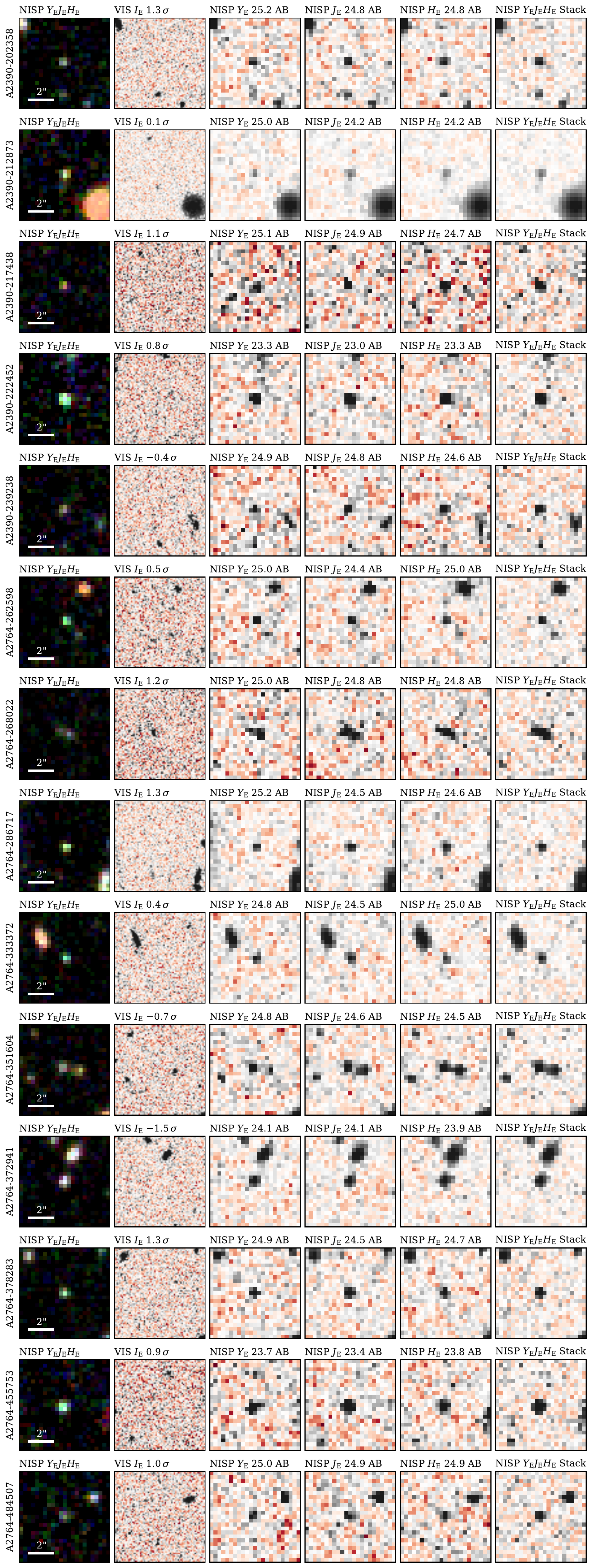}
    \end{subfigure}
    \caption{Cutouts of $z\approx6$--8 Lyman break candidates. Leftmost false-colour RBG images are constructed from NISP/\YJHE. Others show VIS/\IE, NISP/\YJHE, and the NISP detection stack, scaled to $\pm3\sigma$ to emphasize the significance of the detections. Cutouts are 5\arcsec along a side.}
    \label{fig:app:lbg}
\end{figure}

This appendix contains Table~\ref{tab:app:lbg} describing the coordinates, photometry, photometric redshift, and rest-frame UV magnitudes for the 29 LBG candidates, as well as cutouts in Fig.~\ref{fig:app:lbg}.

\begin{table}[H]
\footnotesize\centering
\renewcommand{\arraystretch}{1.35}
\setlength{\tabcolsep}{6pt} 
\begin{threeparttable}
\caption{Coordinates, observed photometry, best-fit $z_{\rm phot}$, and corresponding rest-frame UV magnitude ($M_{\rm UV}$) of the Lyman-break galaxy candidates. \IE photometry is quoted as 1.5\,$\sigma$ lower limits computed from corrected \ang{;;0.6} apertures. \YE, \JE, and \HE photometry is quoted as detections from corrected \ang{;;1.2} apertures. $M_{\rm UV}$ corresponds to the best-fit $z_{\rm phot}$ taken at the peak of $\mathcal{L}(z)$; these are estimated directly from \JE, and are uncorrected for lensing magnification.}
\label{tab:app:lbg}
\begin{tabular}{cccccccccc}
\hline
ID & RA & Dec & Field & \IE & \YE & \JE & \HE & $z_{\rm phot}$ & $M_{\rm UV}$ \\
 & [J2000] & [J2000] &  & [mag] & [mag] & [mag] & [mag]  & & [mag] \\
 \hline
13405 & 328.342108 & 17.264356 & A2390 & 27.45 & $25.30\pm0.31$ & $24.86\pm0.17$ & $25.04\pm0.19$ & 6.46 & $-22.14\pm0.16$ \\
19040 & 328.605616 & 17.299342 & A2390 & 29.48 & $24.61\pm0.27$ & $24.97\pm0.31$ & $24.44\pm0.18$ & 6.69 & $-22.59\pm0.15$ \\
41598 & 328.143886 & 17.405425 & A2390 & 27.63 & $25.51\pm0.35$ & $25.09\pm0.19$ & $25.11\pm0.19$ & 6.46 & $-22.05\pm0.17$ \\
50473 & 328.68049 & 17.44325 & A2390 & 27.90 & $25.23\pm0.26$ & $24.96\pm0.17$ & $25.15\pm0.19$ & 6.54 & $-22.03\pm0.17$ \\
58657 & 328.740945 & 17.474326 & A2390 & 27.55 & $24.21\pm0.09$ & $23.84\pm0.05$ & $24.64\pm0.11$ & 6.61 & $-22.77\pm0.06$ \\
62727 & 328.129815 & 17.488942 & A2390 & 27.48 & $24.39\pm0.11$ & $24.03\pm0.07$ & $24.69\pm0.12$ & 7.17 & $-22.72\pm0.08$ \\
81797 & 328.639259 & 17.552634 & A2390 & 27.48 & $25.24\pm0.23$ & $24.71\pm0.11$ & $25.31\pm0.19$ & 7.67 & $-22.19\pm0.12$ \\
93626 & 328.136037 & 17.589922 & A2390 & 28.79 & $25.03\pm0.18$ & $24.91\pm0.13$ & $24.99\pm0.13$ & 7.00 & $-22.23\pm0.27$ \\
94559 & 328.711201 & 17.592819 & A2390 & 30.29 & $24.80\pm0.17$ & $24.39\pm0.09$ & $24.58\pm0.11$ & 7.67 & $-22.75\pm0.16$ \\
120148 & 327.965859 & 17.674409 & A2390 & 27.24 & $24.89\pm0.25$ & $24.86\pm0.20$ & $25.21\pm0.26$ & 6.61 & $-21.98\pm0.14$ \\
148368 & 327.98274 & 17.763439 & A2390 & 27.43 & $24.60\pm0.13$ & $24.96\pm0.15$ & $24.66\pm0.11$ & 6.46 & $-22.35\pm0.09$ \\
164665 & 328.719457 & 17.816599 & A2390 & 28.27 & $24.88\pm0.19$ & $24.33\pm0.10$ & $24.30\pm0.09$ & 7.17 & $-22.97\pm0.08$ \\
184757 & 328.023821 & 17.885509 & A2390 & 28.84 & $24.51\pm0.12$ & $24.67\pm0.11$ & $25.10\pm0.16$ & 6.85 & $-22.21\pm0.09$ \\
191297 & 328.179802 & 17.909668 & A2390 & 27.94 & $25.64\pm0.32$ & $24.98\pm0.15$ & $25.11\pm0.15$ & 7.85 & $-22.31\pm0.33$ \\
191552 & 328.470833 & 17.910763 & A2390 & 28.21 & $25.18\pm0.22$ & $25.17\pm0.18$ & $25.02\pm0.15$ & 6.61 & $-22.09\pm0.12$ \\
202358 & 328.553214 & 17.953689 & A2390 & 27.67 & $25.41\pm0.28$ & $25.11\pm0.17$ & $25.15\pm0.17$ & 6.46 & $-21.97\pm0.15$ \\
212873 & 328.667034 & 17.997065 & A2390 & 27.98 & $25.21\pm0.31$ & $24.50\pm0.13$ & $24.52\pm0.13$ & 7.67 & $-22.89\pm0.28$ \\
217438 & 328.28835 & 18.015891 & A2390 & 27.61 & $25.37\pm0.29$ & $25.20\pm0.20$ & $25.10\pm0.18$ & 6.32 & $-21.93\pm1.07$ \\
222452 & 328.648599 & 18.037469 & A2390 & 27.40 & $23.52\pm0.07$ & $23.28\pm0.05$ & $23.70\pm0.07$ & 7.25 & $-23.60\pm0.06$ \\
239238 & 328.495839 & 18.134716 & A2390 & 28.97 & $25.08\pm0.21$ & $25.04\pm0.17$ & $24.97\pm0.16$ & 6.85 & $-22.17\pm0.15$ \\
262598 & 5.804307 & $-$49.705047 & A2764 & 28.10 & $25.24\pm0.23$ & $24.63\pm0.11$ & $25.32\pm0.20$ & 7.67 & $-22.24\pm0.12$ \\
268022 & 5.668767 & $-$49.663003 & A2764 & 27.69 & $25.18\pm0.17$ & $25.11\pm0.13$ & $25.20\pm0.14$ & 6.39 & $-21.93\pm0.23$ \\
286717 & 5.546056 & $-$49.561428 & A2764 & 27.80 & $25.42\pm0.23$ & $24.81\pm0.11$ & $24.96\pm0.12$ & 7.76 & $-22.43\pm0.24$ \\
333372 & 6.024066 & $-$49.390368 & A2764 & 28.35 & $25.01\pm0.16$ & $24.82\pm0.11$ & $25.35\pm0.18$ & 6.69 & $-21.95\pm0.09$ \\
351604 & 5.260838 & $-$49.334863 & A2764 & 30.03 & $25.01\pm0.15$ & $24.83\pm0.11$ & $24.84\pm0.11$ & 7.00 & $-22.37\pm0.11$ \\
372941 & 6.34368 & $-$49.272889 & A2764 & 23.90 & $24.28\pm0.08$ & $24.40\pm0.07$ & $24.28\pm0.06$ & 6.85 & $-22.84\pm0.07$ \\
378283 & 5.768291 & $-$49.25985 & A2764 & 27.92 & $25.09\pm0.17$ & $24.73\pm0.10$ & $25.06\pm0.14$ & 6.54 & $-22.16\pm0.12$ \\
455753 & 5.277561 & $-$49.006585 & A2764 & 27.79 & $23.95\pm0.08$ & $23.72\pm0.05$ & $24.13\pm0.08$ & 7.25 & $-23.18\pm0.07$ \\
484507 & 5.536343 & $-$48.856873 & A2764 & 27.78 & $25.18\pm0.25$ & $25.14\pm0.20$ & $25.24\pm0.22$ & 6.46 & $-21.90\pm0.29$ \\
\hline
\end{tabular}
\end{threeparttable}
\end{table}

\pagebreak

\section{Extremely red sources}
\label{app:ero_cutouts}
This appendix contains Tables~\ref{tab:app:ers_1}, \ref{tab:app:ers_2}, \ref{tab:app:ers_3} and \ref{tab:app:ers_4} describing the coordinates, photometry, and photometric redshift solution(s) for the 139 ERSs, as well as cutouts in Figs.~\ref{fig:app:ers_1}, \ref{fig:app:ers_2}, \ref{fig:app:ers_3} and \ref{fig:app:ers_4}.

\begin{table}[H]
\footnotesize\centering
\renewcommand{\arraystretch}{1.35}
\setlength{\tabcolsep}{6pt} 
\begin{threeparttable}
\caption{Coordinates, observed photometry, best-fit $z_{\rm phot}$, and corresponding redshfit solution(s) of the Extremely Red sources. \IE photometry is quoted as 1.5\,$\sigma$ upper limits computed from corrected \ang{;;0.6} apertures. \YE, \JE, \HE photometry is quoted as detections from corrected \ang{;;1.2} apertures. The best-fit photometric redshfit $z_{\rm phot}$ quoted here is taken at the peak of $\mathcal{L}(z>5)$, with any significant peak at $z<5$ indicated for bimodal solutions.}
\label{tab:app:ers_1}
\begin{tabular}{cccccccccc}
\hline
ID & RA & Dec & Field & \IE & \YE & \JE & \HE & $z_{\rm phot}$ & $z_{\rm phot}(<5)$  \\
 & [J2000] & [J2000] &  & [mag] & [mag] & [mag] & [mag]  & &  \\
 \hline
1418 & 328.256848 & 17.153557 & A2390 & 26.61 & $24.97\pm0.32$ & $24.15\pm0.12$ & $23.50\pm0.06$ & 5.69 & 1.44 \\
11725 & 328.392774 & 17.251623 & A2390 & 27.22 & $25.15\pm0.29$ & $24.51\pm0.13$ & $23.55\pm0.05$ & 5.42 & 1.41 \\
12491 & 328.192359 & 17.257306 & A2390 & 27.17 & $25.14\pm0.32$ & $24.88\pm0.21$ & $24.56\pm0.15$ & 6.32 & 1.46 \\
17513 & 328.197604 & 17.290342 & A2390 & 27.31 & $25.18\pm0.25$ & $24.63\pm0.13$ & $23.75\pm0.05$ & 5.62 & 4.99 \\
20071 & 328.443517 & 17.304954 & A2390 & 27.38 & $24.72\pm0.17$ & $23.79\pm0.06$ & $23.06\pm0.03$ & 5.69 & 1.56 \\
21171 & 328.51353 & 17.310914 & A2390 & 27.56 & $25.00\pm0.29$ & $24.77\pm0.19$ & $24.12\pm0.10$ & 5.89 & 1.46 \\
29808 & 328.307984 & 17.355299 & A2390 & 27.36 & $25.49\pm0.29$ & $24.76\pm0.12$ & $24.24\pm0.07$ & 5.69 & 1.46 \\
32789 & 328.354038 & 17.369211 & A2390 & 30.90 & $25.10\pm0.19$ & $24.92\pm0.13$ & $24.50\pm0.08$ & 6.69 & -- \\
33926 & 328.586033 & 17.373971 & A2390 & 27.92 & $25.46\pm0.34$ & $24.46\pm0.11$ & $23.92\pm0.07$ & 6.54 & 1.64 \\
34754 & 328.338724 & 17.377432 & A2390 & 27.49 & $25.00\pm0.19$ & $24.64\pm0.11$ & $23.97\pm0.06$ & 5.82 & 1.44 \\
42274 & 328.32914 & 17.408282 & A2390 & 28.78 & $25.51\pm0.30$ & $25.19\pm0.18$ & $24.48\pm0.09$ & 6.10 & 1.51 \\
44724 & 328.228369 & 17.418534 & A2390 & 28.27 & $24.53\pm0.12$ & $23.92\pm0.06$ & $23.30\pm0.03$ & 6.24 & -- \\
46207 & 328.627413 & 17.425007 & A2390 & 28.06 & $25.66\pm0.34$ & $24.56\pm0.11$ & $23.95\pm0.06$ & 6.69 & 1.72 \\
48955 & 328.365466 & 17.436572 & A2390 & 28.03 & $25.15\pm0.23$ & $24.13\pm0.07$ & $23.58\pm0.04$ & 6.54 & -- \\
49401 & 328.311951 & 17.438465 & A2390 & 27.61 & $24.78\pm0.16$ & $23.70\pm0.05$ & $22.87\pm0.02$ & 5.69 & 1.56 \\
52137 & 328.167879 & 17.443979 & A2390 & 27.86 & $25.57\pm0.31$ & $24.27\pm0.08$ & $23.59\pm0.04$ & 7.00 & 1.83 \\
52813 & 328.482966 & 17.45313 & A2390 & 28.00 & $25.61\pm0.31$ & $25.15\pm0.17$ & $24.67\pm0.10$ & 6.10 & 1.46 \\
57400 & 328.673108 & 17.469908 & A2390 & 27.71 & $24.88\pm0.17$ & $24.26\pm0.08$ & $23.45\pm0.04$ & 5.89 & 1.51 \\
61030 & 328.328541 & 17.483197 & A2390 & 28.05 & $25.15\pm0.22$ & $24.56\pm0.10$ & $23.53\pm0.04$ & 5.69 & 1.39 \\
62830 & 328.654118 & 17.489255 & A2390 & 29.31 & $24.63\pm0.14$ & $23.60\pm0.04$ & $22.83\pm0.02$ & 6.46 & -- \\
72934 & 328.209521 & 17.524068 & A2390 & 28.32 & $25.68\pm0.34$ & $24.59\pm0.10$ & $23.92\pm0.05$ & 6.61 & 1.80 \\
75674 & 328.335908 & 17.532969 & A2390 & 27.54 & $25.72\pm0.34$ & $24.42\pm0.09$ & $23.77\pm0.04$ & 6.92 & 1.92 \\
86384 & 328.461045 & 17.567449 & A2390 & 28.22 & $25.66\pm0.36$ & $24.56\pm0.11$ & $23.69\pm0.05$ & 5.69 & 1.61 \\
90220 & 328.816407 & 17.579096 & A2390 & 27.67 & $24.94\pm0.18$ & $24.02\pm0.06$ & $23.36\pm0.03$ & 6.39 & 1.64 \\
90763 & 328.4943 & 17.581134 & A2390 & 27.72 & $25.50\pm0.28$ & $24.36\pm0.08$ & $23.71\pm0.04$ & 6.69 & 1.75 \\
90795 & 328.787715 & 17.580962 & A2390 & 28.10 & $25.69\pm0.33$ & $24.69\pm0.10$ & $24.14\pm0.06$ & 6.54 & 1.64 \\
93645 & 327.999493 & 17.589824 & A2390 & 26.57 & $25.24\pm0.35$ & $24.72\pm0.18$ & $24.10\pm0.10$ & 5.55 & 1.41 \\
94124 & 328.048253 & 17.591283 & A2390 & 27.47 & $25.45\pm0.33$ & $24.13\pm0.08$ & $22.95\pm0.03$ & 5.69 & 1.56 \\
94349 & 328.046741 & 17.591861 & A2390 & 27.08 & $24.67\pm0.16$ & $23.79\pm0.06$ & $22.97\pm0.03$ & 5.69 & 1.51 \\
\hline
\end{tabular}
\end{threeparttable}
\end{table}

\begin{table*}[t!]
\footnotesize\centering
\renewcommand{\arraystretch}{1.35}
\setlength{\tabcolsep}{6pt} 
\begin{threeparttable}
\caption{Continued...}
\label{tab:app:ers_2}
\begin{tabular}{cccccccccc}
\hline
ID & RA & Dec & Field & \IE & \YE & \JE & \HE & $z_{\rm phot}$ & $z_{\rm phot}(<5)$  \\
 & [J2000] & [J2000] &  & [mag] & [mag] & [mag] & [mag]  & &  \\
 \hline
96461 & 328.061955 & 17.599076 & A2390 & 28.65 & $25.34\pm0.27$ & $24.64\pm0.12$ & $24.05\pm0.07$ & 6.39 & 1.56 \\
97193 & 328.330676 & 17.601779 & A2390 & 28.04 & $25.67\pm0.33$ & $24.71\pm0.11$ & $23.84\pm0.05$ & 5.69 & 1.46 \\
98610 & 328.026642 & 17.606334 & A2390 & 27.92 & $25.19\pm0.26$ & $24.22\pm0.09$ & $23.49\pm0.04$ & 6.32 & 1.59 \\
99206 & 328.047244 & 17.608267 & A2390 & 27.79 & $25.59\pm0.35$ & $24.79\pm0.14$ & $24.17\pm0.07$ & 6.39 & 1.56 \\
100728 & 328.164931 & 17.613177 & A2390 & 27.91 & $24.88\pm0.18$ & $24.47\pm0.10$ & $23.58\pm0.04$ & 5.69 & -- \\
102521 & 328.087851 & 17.618922 & A2390 & 27.34 & $25.51\pm0.27$ & $24.88\pm0.13$ & $24.48\pm0.09$ & 5.69 & 1.46 \\
106154 & 328.294392 & 17.630897 & A2390 & 27.54 & $25.05\pm0.19$ & $24.25\pm0.07$ & $23.49\pm0.03$ & 5.69 & 1.53 \\
110635 & 328.360005 & 17.644253 & A2390 & 28.03 & $25.50\pm0.30$ & $24.33\pm0.08$ & $23.72\pm0.05$ & 6.77 & 1.77 \\
116975 & 328.380004 & 17.664733 & A2390 & 28.23 & $25.39\pm0.26$ & $24.83\pm0.13$ & $24.43\pm0.08$ & 6.32 & -- \\
117303 & 328.797486 & 17.665343 & A2390 & 27.61 & $24.70\pm0.18$ & $24.07\pm0.08$ & $23.87\pm0.06$ & 6.46 & -- \\
123765 & 328.258542 & 17.685761 & A2390 & 27.87 & $25.38\pm0.25$ & $24.63\pm0.10$ & $23.79\pm0.05$ & 5.69 & 1.51 \\
126542 & 328.00841 & 17.694133 & A2390 & 27.16 & $25.46\pm0.31$ & $24.56\pm0.12$ & $23.63\pm0.05$ & 5.69 & 1.41 \\
126907 & 328.736884 & 17.695487 & A2390 & 27.85 & $25.55\pm0.32$ & $24.69\pm0.12$ & $23.96\pm0.06$ & 5.69 & 1.53 \\
127591 & 328.332861 & 17.697974 & A2390 & 27.70 & $24.92\pm0.18$ & $24.50\pm0.10$ & $23.73\pm0.05$ & 5.89 & -- \\
127981 & 328.577841 & 17.699035 & A2390 & 27.52 & $25.58\pm0.30$ & $24.53\pm0.10$ & $23.63\pm0.04$ & 5.69 & 1.39 \\
128262 & 328.131084 & 17.699773 & A2390 & 27.87 & $25.10\pm0.20$ & $24.34\pm0.08$ & $23.76\pm0.04$ & 6.39 & -- \\
128275 & 328.649621 & 17.699791 & A2390 & 27.68 & $24.96\pm0.17$ & $24.46\pm0.09$ & $23.65\pm0.04$ & 5.82 & 1.51 \\
130298 & 327.986255 & 17.70546 & A2390 & 27.15 & $25.13\pm0.26$ & $24.27\pm0.09$ & $23.82\pm0.06$ & 5.69 & 1.61 \\
130523 & 328.510018 & 17.706703 & A2390 & 27.65 & $25.37\pm0.26$ & $24.60\pm0.10$ & $24.13\pm0.06$ & 5.69 & 1.46 \\
131192 & 328.655842 & 17.708628 & A2390 & 23.90 & $25.63\pm0.35$ & $25.21\pm0.20$ & $24.29\pm0.08$ & 6.10 & 1.46 \\
132329 & 328.566782 & 17.712264 & A2390 & 27.88 & $25.69\pm0.32$ & $25.10\pm0.15$ & $24.50\pm0.08$ & 5.62 & 1.48 \\
132673 & 328.066394 & 17.71305 & A2390 & 27.59 & $25.50\pm0.28$ & $24.90\pm0.13$ & $24.21\pm0.07$ & 5.55 & 1.48 \\
132838 & 327.951613 & 17.713357 & A2390 & 28.04 & $25.21\pm0.30$ & $24.37\pm0.11$ & $23.83\pm0.07$ & 6.54 & 1.56 \\
134238 & 328.710895 & 17.71771 & A2390 & 27.89 & $25.32\pm0.23$ & $24.64\pm0.10$ & $23.86\pm0.05$ & 5.96 & 1.51 \\
134239 & 328.627428 & 17.71786 & A2390 & 27.47 & $25.65\pm0.32$ & $25.05\pm0.15$ & $24.56\pm0.09$ & 5.69 & 1.46 \\
134611 & 327.928143 & 17.718602 & A2390 & 26.51 & $25.28\pm0.35$ & $23.82\pm0.08$ & $23.25\pm0.04$ & 7.17 & 2.06 \\
135122 & 328.587476 & 17.720894 & A2390 & 27.71 & $25.63\pm0.35$ & $25.01\pm0.16$ & $24.63\pm0.11$ & 5.69 & 1.46 \\
135949 & 327.958728 & 17.723145 & A2390 & 27.07 & $24.96\pm0.21$ & $23.57\pm0.05$ & $22.63\pm0.02$ & 5.69 & 1.75 \\
136075 & 328.570979 & 17.724072 & A2390 & 27.57 & $24.75\pm0.14$ & $23.74\pm0.05$ & $22.87\pm0.02$ & 5.69 & 1.46 \\
139572 & 327.976023 & 17.734667 & A2390 & 27.00 & $25.47\pm0.32$ & $25.03\pm0.17$ & $24.14\pm0.07$ & 5.05 & 4.99 \\
140449 & 328.108724 & 17.737865 & A2390 & 27.68 & $25.30\pm0.25$ & $25.01\pm0.16$ & $24.77\pm0.12$ & 6.39 & -- \\
140700 & 328.594794 & 17.738527 & A2390 & 27.83 & $24.68\pm0.14$ & $23.77\pm0.05$ & $23.03\pm0.02$ & 6.24 & -- \\
142039 & 328.646716 & 17.742988 & A2390 & 28.00 & $25.13\pm0.19$ & $24.87\pm0.12$ & $24.06\pm0.05$ & 5.82 & -- \\
146825 & 328.014082 & 17.75839 & A2390 & 28.23 & $25.38\pm0.25$ & $24.83\pm0.12$ & $24.15\pm0.06$ & 6.10 & 1.51 \\
148129 & 328.622533 & 17.762907 & A2390 & 27.50 & $25.59\pm0.32$ & $24.46\pm0.09$ & $23.93\pm0.05$ & 6.92 & 1.83 \\
152601 & 327.947827 & 17.776723 & A2390 & 27.42 & $25.26\pm0.24$ & $24.87\pm0.14$ & $23.84\pm0.05$ & 5.69 & 4.99 \\
156298 & 328.268199 & 17.788413 & A2390 & 28.39 & $25.32\pm0.27$ & $24.79\pm0.13$ & $23.86\pm0.05$ & 5.82 & 1.44 \\
156920 & 328.643826 & 17.791403 & A2390 & 27.83 & $25.72\pm0.34$ & $24.67\pm0.11$ & $24.10\pm0.06$ & 6.46 & 1.64 \\
159130 & 328.576532 & 17.798692 & A2390 & 28.16 & $25.69\pm0.33$ & $24.64\pm0.11$ & $23.83\pm0.05$ & 5.69 & 1.56 \\
160963 & 328.411422 & 17.804849 & A2390 & 27.69 & $25.54\pm0.30$ & $24.86\pm0.13$ & $23.79\pm0.05$ & 5.36 & 4.99 \\
160990 & 328.428243 & 17.804948 & A2390 & 28.04 & $25.53\pm0.30$ & $25.06\pm0.16$ & $24.50\pm0.09$ & 6.10 & 1.46 \\
162124 & 328.003353 & 17.808232 & A2390 & 27.69 & $25.37\pm0.23$ & $24.89\pm0.13$ & $24.67\pm0.10$ & 6.03 & -- \\
174313 & 328.114589 & 17.849632 & A2390 & 27.52 & $25.61\pm0.31$ & $24.92\pm0.13$ & $24.18\pm0.06$ & 5.69 & 1.39 \\
176069 & 328.546806 & 17.855855 & A2390 & 27.93 & $25.50\pm0.28$ & $24.85\pm0.13$ & $23.96\pm0.06$ & 5.69 & 1.44 \\
178357 & 328.58747 & 17.863986 & A2390 & 28.36 & $25.04\pm0.19$ & $24.09\pm0.06$ & $23.35\pm0.03$ & 6.32 & -- \\
\hline
\end{tabular}

\end{threeparttable}
\end{table*}

\begin{table*}[t!]
\footnotesize\centering
\renewcommand{\arraystretch}{1.35}
\setlength{\tabcolsep}{6pt} 
\begin{threeparttable}
\caption{Continued...}
\label{tab:app:ers_3}
\begin{tabular}{cccccccccc}
\hline
ID & RA & Dec & Field & \IE & \YE & \JE & \HE & $z_{\rm phot}$ & $z_{\rm phot}(<5)$  \\
 & [J2000] & [J2000] &  & [mag] & [mag] & [mag] & [mag]  & &  \\
 \hline
179032 & 328.513208 & 17.866424 & A2390 & 28.42 & $25.53\pm0.33$ & $25.02\pm0.16$ & $24.76\pm0.13$ & 7.00 & -- \\
181995 & 328.272992 & 17.876456 & A2390 & 27.49 & $25.18\pm0.22$ & $24.56\pm0.10$ & $23.61\pm0.04$ & 5.42 & 1.36 \\
182016 & 328.623151 & 17.875228 & A2390 & 28.01 & $23.97\pm0.07$ & $23.47\pm0.04$ & $22.89\pm0.02$ & 6.24 & -- \\
182511 & 328.608852 & 17.878096 & A2390 & 29.33 & $25.52\pm0.26$ & $25.21\pm0.17$ & $24.97\pm0.12$ & 6.85 & -- \\
184713 & 328.066672 & 17.884999 & A2390 & 27.73 & $24.89\pm0.15$ & $24.38\pm0.08$ & $24.01\pm0.05$ & 6.03 & -- \\
186655 & 328.063303 & 17.892486 & A2390 & 27.81 & $25.50\pm0.27$ & $24.48\pm0.09$ & $23.54\pm0.04$ & 5.69 & 1.44 \\
190003 & 328.452507 & 17.904887 & A2390 & 27.52 & $25.26\pm0.24$ & $24.26\pm0.08$ & $23.57\pm0.04$ & 5.69 & 1.66 \\
190359 & 328.194595 & 17.906096 & A2390 & 28.34 & $25.57\pm0.29$ & $24.77\pm0.11$ & $23.79\pm0.04$ & 5.69 & 1.39 \\
198968 & 328.498325 & 17.939391 & A2390 & 28.03 & $25.57\pm0.32$ & $24.39\pm0.09$ & $23.56\pm0.04$ & 5.69 & 1.61 \\
200432 & 328.612065 & 17.945323 & A2390 & 27.55 & $25.52\pm0.31$ & $24.53\pm0.10$ & $23.79\pm0.05$ & 5.69 & 1.56 \\
200510 & 328.695125 & 17.945585 & A2390 & 27.21 & $24.88\pm0.25$ & $24.42\pm0.14$ & $24.12\pm0.10$ & 5.89 & -- \\
201258 & 328.459279 & 17.948958 & A2390 & 28.08 & $25.51\pm0.32$ & $24.99\pm0.16$ & $24.51\pm0.10$ & 6.17 & 1.46 \\
213211 & 328.052488 & 17.998072 & A2390 & 27.01 & $24.55\pm0.21$ & $24.10\pm0.11$ & $23.29\pm0.05$ & 5.75 & 1.39 \\
217816 & 328.675502 & 18.017266 & A2390 & 27.23 & $24.61\pm0.20$ & $23.91\pm0.09$ & $23.73\pm0.07$ & 6.46 & -- \\
227487 & 328.314319 & 18.063201 & A2390 & 27.63 & $25.08\pm0.22$ & $24.56\pm0.11$ & $24.18\pm0.08$ & 5.96 & -- \\
255757 & 5.69298 & $-$49.77244 & A2764 & 27.26 & $25.43\pm0.29$ & $25.25\pm0.21$ & $24.45\pm0.10$ & 5.62 & 4.99 \\
258762 & 5.761004 & $-$49.740345 & A2764 & 27.63 & $25.69\pm0.34$ & $25.05\pm0.16$ & $24.11\pm0.07$ & 5.42 & 1.41 \\
260052 & 5.737694 & $-$49.727595 & A2764 & 27.93 & $24.45\pm0.10$ & $23.85\pm0.05$ & $23.01\pm0.02$ & 5.96 & -- \\
275980 & 5.910008 & $-$49.614452 & A2764 & 27.68 & $25.08\pm0.18$ & $24.10\pm0.06$ & $23.22\pm0.03$ & 5.69 & 1.46 \\
276361 & 5.623303 & $-$49.61231 & A2764 & 27.93 & $25.38\pm0.21$ & $25.02\pm0.12$ & $24.28\pm0.06$ & 5.82 & 1.44 \\
277056 & 5.591612 & $-$49.608529 & A2764 & 27.64 & $25.59\pm0.28$ & $24.76\pm0.11$ & $24.09\pm0.06$ & 5.69 & 1.53 \\
277236 & 5.504416 & $-$49.607464 & A2764 & 28.22 & $25.50\pm0.29$ & $25.20\pm0.19$ & $24.17\pm0.07$ & 5.69 & 4.99 \\
279750 & 5.565212 & $-$49.594306 & A2764 & 27.75 & $25.67\pm0.27$ & $24.94\pm0.12$ & $24.35\pm0.07$ & 5.69 & 1.51 \\
290229 & 5.931895 & $-$49.546567 & A2764 & 27.81 & $25.59\pm0.27$ & $24.53\pm0.08$ & $23.84\pm0.04$ & 5.69 & 1.69 \\
297920 & 5.346424 & $-$49.514259 & A2764 & 27.89 & $25.61\pm0.32$ & $24.83\pm0.13$ & $24.02\pm0.06$ & 5.69 & 1.51 \\
300371 & 5.429171 & $-$49.50444 & A2764 & 27.96 & $25.62\pm0.27$ & $24.90\pm0.12$ & $24.14\pm0.06$ & 5.62 & 1.51 \\
307317 & 5.5597 & $-$49.478114 & A2764 & 27.99 & $25.06\pm0.17$ & $24.22\pm0.07$ & $23.60\pm0.04$ & 6.39 & -- \\
311115 & 6.063442 & $-$49.463984 & A2764 & 29.53 & $25.04\pm0.16$ & $24.47\pm0.08$ & $23.95\pm0.05$ & 6.46 & -- \\
312245 & 5.758472 & $-$49.460722 & A2764 & 28.04 & $25.21\pm0.18$ & $24.27\pm0.06$ & $23.53\pm0.03$ & 5.69 & 1.56 \\
316484 & 6.046452 & $-$49.444592 & A2764 & 27.95 & $25.32\pm0.21$ & $24.28\pm0.07$ & $23.49\pm0.03$ & 5.69 & 1.53 \\
316921 & 5.318509 & $-$49.443448 & A2764 & 27.61 & $25.66\pm0.30$ & $24.53\pm0.09$ & $23.78\pm0.04$ & 5.69 & 1.69 \\
317466 & 5.96933 & $-$49.441346 & A2764 & 28.47 & $25.66\pm0.27$ & $25.12\pm0.14$ & $24.29\pm0.06$ & 5.89 & 1.51 \\
318092 & 6.127517 & $-$49.439122 & A2764 & 28.83 & $25.58\pm0.25$ & $24.59\pm0.09$ & $24.18\pm0.06$ & 7.17 & -- \\
331934 & 5.209565 & $-$49.394328 & A2764 & 27.38 & $25.12\pm0.20$ & $25.07\pm0.16$ & $24.20\pm0.07$ & 5.69 & 4.99 \\
339812 & 5.754469 & $-$49.37093 & A2764 & 28.05 & $25.45\pm0.25$ & $24.25\pm0.07$ & $23.46\pm0.03$ & 5.69 & 1.69 \\
342320 & 6.397808 & $-$49.361065 & A2764 & 27.73 & $25.36\pm0.27$ & $24.91\pm0.14$ & $24.34\pm0.08$ & 5.89 & 1.46 \\
342359 & 5.594825 & $-$49.363044 & A2764 & 28.64 & $25.56\pm0.25$ & $23.98\pm0.05$ & $23.17\pm0.02$ & 7.17 & 1.80 \\
349646 & 6.013694 & $-$49.340953 & A2764 & 27.89 & $25.36\pm0.21$ & $25.15\pm0.14$ & $24.75\pm0.10$ & 6.39 & -- \\
351918 & 6.383198 & $-$49.332837 & A2764 & 27.99 & $25.57\pm0.28$ & $25.15\pm0.15$ & $24.82\pm0.11$ & 6.03 & 1.46 \\
352361 & 6.40128 & $-$49.3315 & A2764 & 27.85 & $25.71\pm0.33$ & $24.94\pm0.13$ & $24.14\pm0.06$ & 5.69 & 1.51 \\
359339 & 5.498303 & $-$49.313059 & A2764 & 28.29 & $25.59\pm0.26$ & $25.25\pm0.16$ & $24.58\pm0.09$ & 5.96 & 1.51 \\
361593 & 5.665488 & $-$49.306737 & A2764 & 28.93 & $25.61\pm0.31$ & $24.99\pm0.15$ & $24.36\pm0.08$ & 6.32 & 1.56 \\
368096 & 5.982153 & $-$49.287806 & A2764 & 28.94 & $24.98\pm0.14$ & $24.32\pm0.07$ & $23.45\pm0.03$ & 5.96 & -- \\
368343 & 6.110498 & $-$49.28676 & A2764 & 28.15 & $25.52\pm0.27$ & $25.04\pm0.14$ & $24.06\pm0.06$ & 5.69 & 4.99 \\
383133 & 5.998861 & $-$49.246759 & A2764 & 27.97 & $25.63\pm0.27$ & $25.07\pm0.14$ & $24.30\pm0.07$ & 5.62 & 1.51 \\

\hline
\end{tabular}
\end{threeparttable}
\end{table*}

\begin{table*}[t!]
\footnotesize\centering
\renewcommand{\arraystretch}{1.35}
\setlength{\tabcolsep}{6pt} 
\begin{threeparttable}
\caption{Continued...}
\label{tab:app:ers_4}
\begin{tabular}{cccccccccc}
\hline
ID & RA & Dec & Field & \IE & \YE & \JE & \HE & $z_{\rm phot}$ & $z_{\rm phot}(<5)$  \\
 & [J2000] & [J2000] &  & [mag] & [mag] & [mag] & [mag]  & &  \\
 \hline
388197 & 6.015296 & $-$49.232755 & A2764 & 28.02 & $25.22\pm0.19$ & $24.71\pm0.10$ & $23.80\pm0.04$ & 5.69 & 1.39 \\
388530 & 5.711416 & $-$49.232415 & A2764 & 27.89 & $25.36\pm0.20$ & $24.67\pm0.09$ & $23.80\pm0.04$ & 5.62 & 1.44 \\
399947 & 6.271704 & $-$49.197094 & A2764 & 28.30 & $25.54\pm0.33$ & $24.47\pm0.10$ & $23.44\pm0.04$ & 5.69 & 1.46 \\
400268 & 5.034318 & $-$49.195575 & A2764 & 28.08 & $25.51\pm0.27$ & $24.68\pm0.11$ & $23.79\pm0.05$ & 5.69 & 1.46 \\
409974 & 6.358018 & $-$49.16543 & A2764 & 27.23 & $24.87\pm0.22$ & $24.16\pm0.09$ & $23.64\pm0.06$ & 5.69 & 1.46 \\
412279 & 6.074586 & $-$49.159629 & A2764 & 28.75 & $25.39\pm0.21$ & $24.71\pm0.10$ & $23.97\pm0.05$ & 6.10 & -- \\
412992 & 6.182453 & $-$49.157207 & A2764 & 27.94 & $25.69\pm0.29$ & $24.73\pm0.10$ & $24.19\pm0.06$ & 6.39 & 1.59 \\
416453 & 6.134152 & $-$49.147264 & A2764 & 27.83 & $25.34\pm0.21$ & $24.48\pm0.08$ & $23.94\pm0.05$ & 5.69 & 1.56 \\
417316 & 5.000542 & $-$49.143334 & A2764 & 27.62 & $25.32\pm0.32$ & $24.24\pm0.10$ & $23.59\pm0.05$ & 6.54 & 1.75 \\
419641 & 5.32974 & $-$49.137648 & A2764 & 28.36 & $25.52\pm0.23$ & $24.89\pm0.11$ & $24.14\pm0.05$ & 6.03 & 1.53 \\
429295 & 5.306015 & $-$49.104542 & A2764 & 28.12 & $25.55\pm0.25$ & $24.44\pm0.07$ & $23.89\pm0.04$ & 6.92 & 1.69 \\
435610 & 5.503306 & $-$49.083999 & A2764 & 27.88 & $25.50\pm0.24$ & $24.66\pm0.09$ & $23.74\pm0.04$ & 5.69 & 1.44 \\
444019 & 5.850835 & $-$49.053654 & A2764 & 28.38 & $24.81\pm0.14$ & $24.07\pm0.06$ & $23.32\pm0.03$ & 6.10 & -- \\
450433 & 5.935346 & $-$49.028956 & A2764 & 28.00 & $25.68\pm0.26$ & $24.82\pm0.10$ & $24.37\pm0.07$ & 6.32 & 1.53 \\
461554 & 5.355018 & $-$48.979285 & A2764 & 28.02 & $24.93\pm0.19$ & $24.32\pm0.09$ & $23.47\pm0.04$ & 5.89 & 1.51 \\
472651 & 5.910263 & $-$48.926929 & A2764 & 28.29 & $25.50\pm0.29$ & $24.14\pm0.07$ & $23.40\pm0.03$ & 7.08 & 1.92 \\
474209 & 5.306756 & $-$48.918358 & A2764 & 27.90 & $24.75\pm0.26$ & $24.16\pm0.13$ & $23.79\pm0.09$ & 6.85 & -- \\
474659 & 5.359467 & $-$48.915954 & A2764 & 26.89 & $24.97\pm0.33$ & $24.59\pm0.19$ & $24.03\pm0.11$ & 5.69 & 1.39 \\
482426 & 5.400998 & $-$48.883814 & A2764 & 27.35 & $24.39\pm0.19$ & $23.26\pm0.06$ & $22.55\pm0.03$ & 6.61 & 1.69 \\
499104 & 5.756752 & $-$48.719152 & A2764 & 29.29 & $24.93\pm0.26$ & $24.23\pm0.12$ & $23.78\pm0.07$ & 6.85 & -- \\

\hline
\end{tabular}
\end{threeparttable}
\end{table*}

\pagebreak

\begin{figure}%
    \centering
    \begin{subfigure}[t]{0.5\textwidth}
        \centering
        \includegraphics[width=0.92\textwidth]{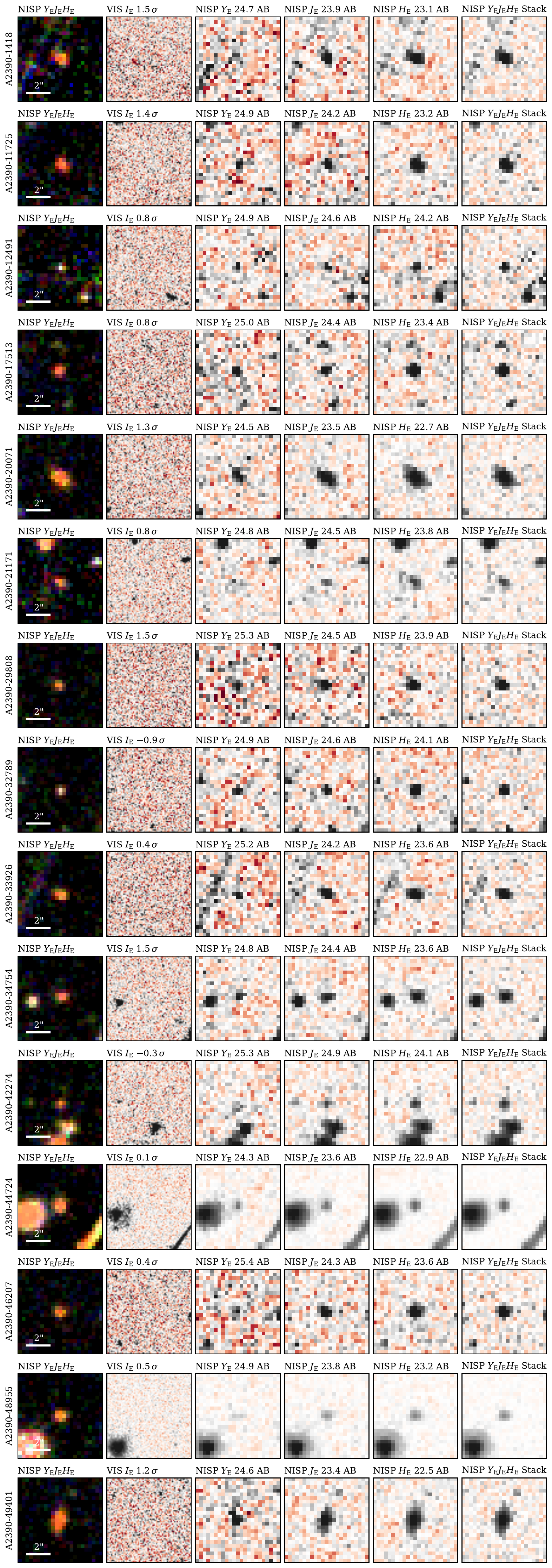}
    \end{subfigure}%
    ~ 
    \begin{subfigure}[t]{0.5\textwidth}
        \centering
        \includegraphics[width=0.92\textwidth]{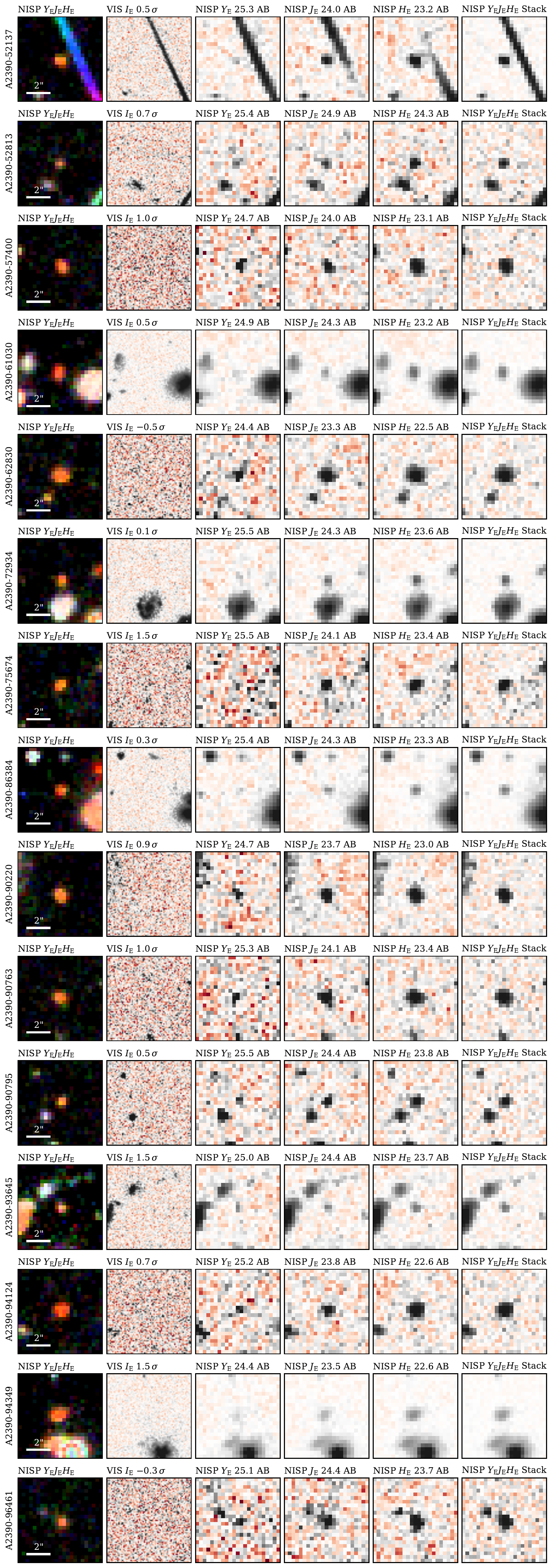}
    \end{subfigure}
    \caption{Cutouts of extremely red sources. Leftmost false-colour RBG images are constructed from NISP/\YJHE. Others show VIS/\IE, NISP/\YJHE, and the NISP detection stack, scaled to $\pm3\sigma$ to emphasize the significance of the detections. Cutouts are 5\arcsec along a side.}
    \label{fig:app:ers_1}
\end{figure}

\begin{figure}%
    \centering
    \begin{subfigure}[T]{0.5\textwidth}
        \centering
        \includegraphics[width=0.92\textwidth]{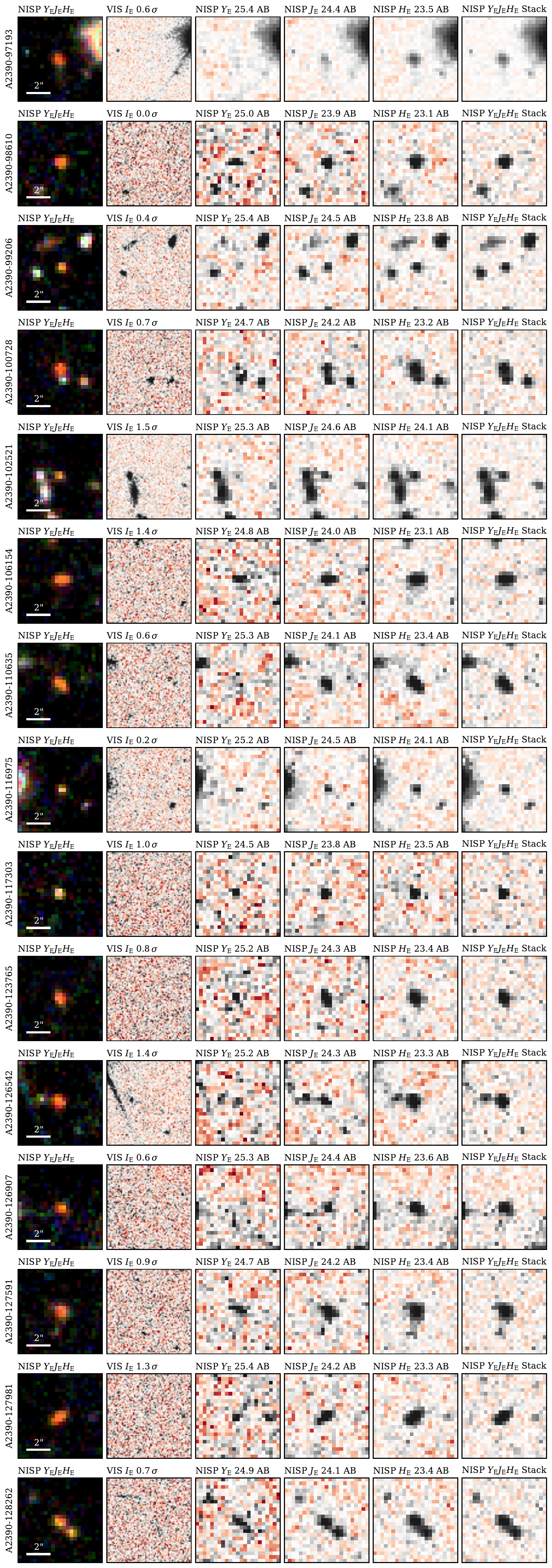}
    \end{subfigure}%
    ~ 
    \begin{subfigure}[T]{0.5\textwidth}
        \centering
        \includegraphics[width=0.92\textwidth]{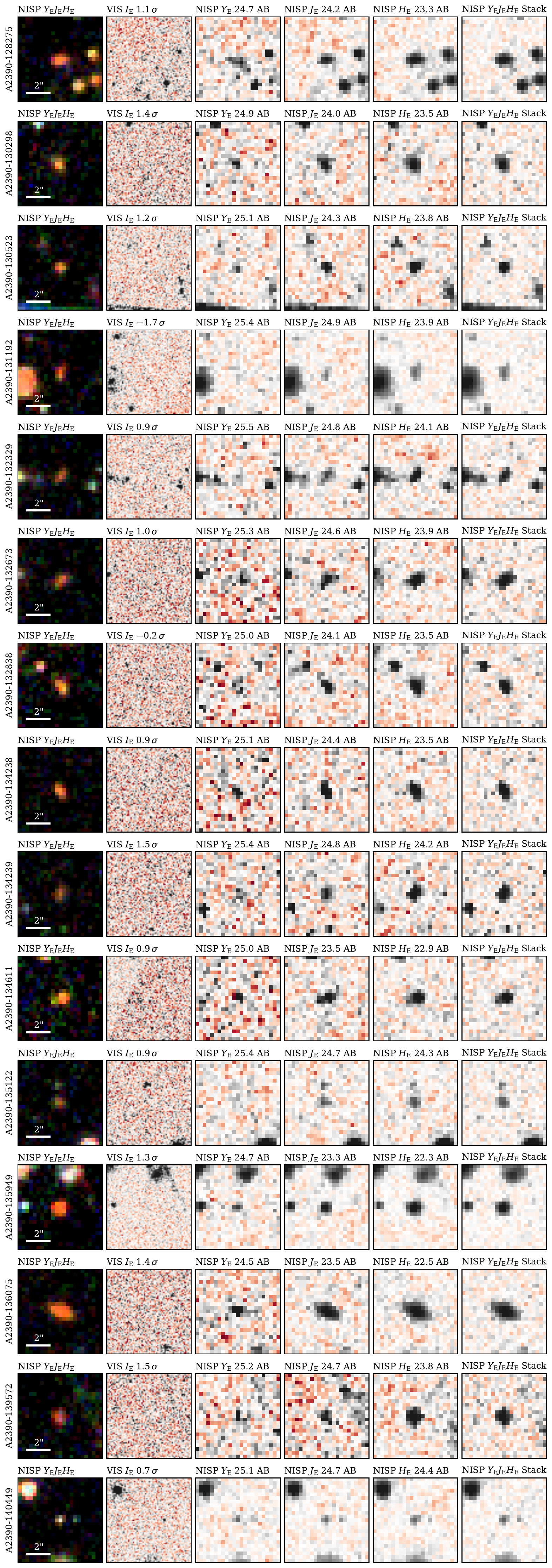}
    \end{subfigure}
    \caption{Continued...}
    \label{fig:app:ers_2}
\end{figure}

\begin{figure}%
    \centering
    \begin{subfigure}[T]{0.5\textwidth}
        \centering
        \includegraphics[width=0.92\textwidth]{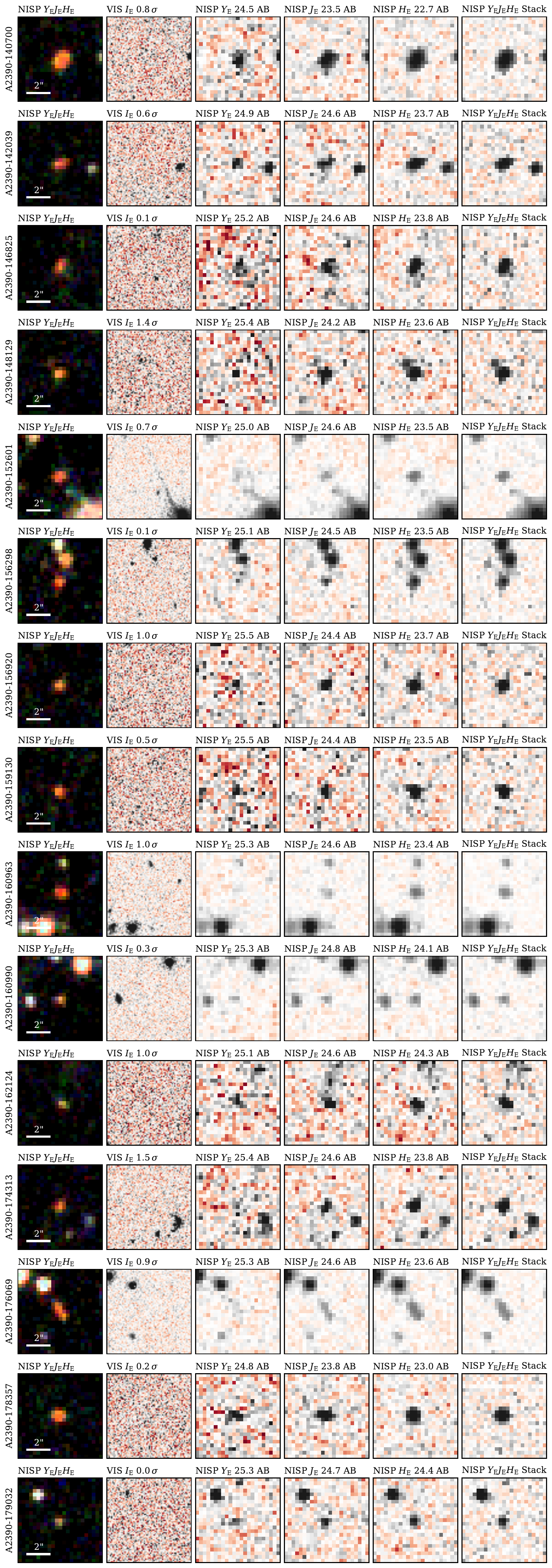}
    \end{subfigure}%
    ~ 
    \begin{subfigure}[T]{0.5\textwidth}
        \centering
        \includegraphics[width=0.92\textwidth]{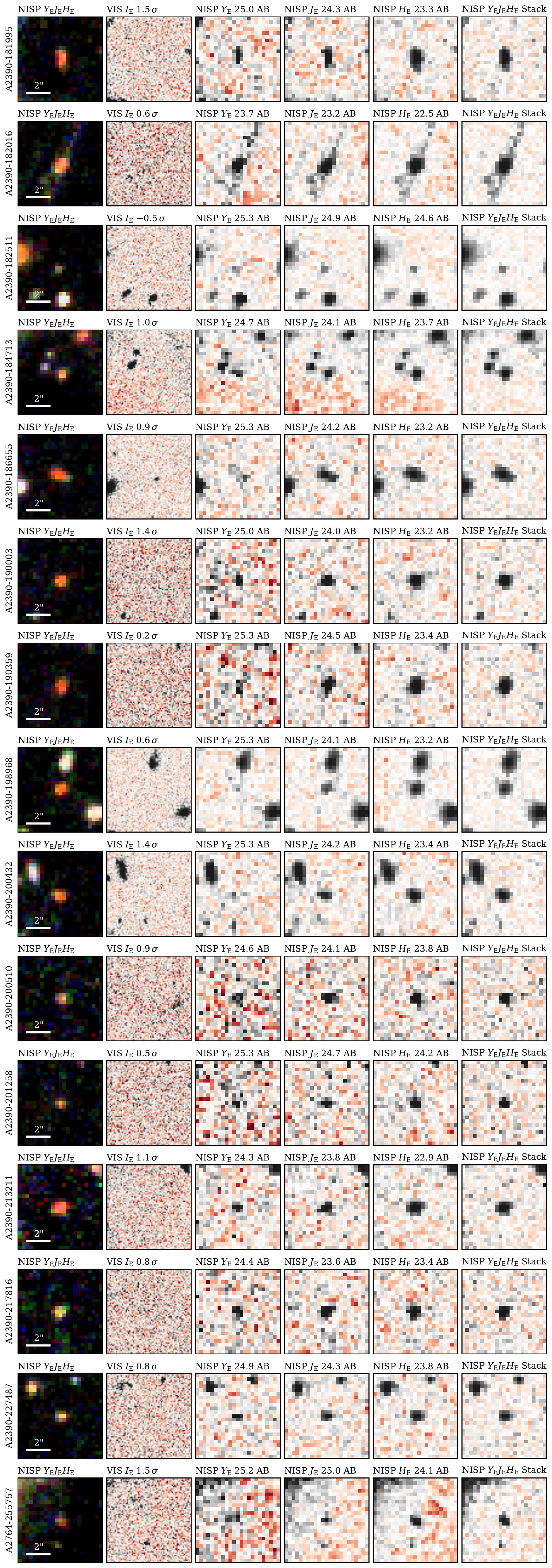}
    \end{subfigure}
    \caption{Continued...}
    \label{fig:app:ers_3}
\end{figure}

\begin{figure}%
    \centering
    \begin{subfigure}[T]{0.5\textwidth}
        \centering
        \includegraphics[width=0.92\textwidth]{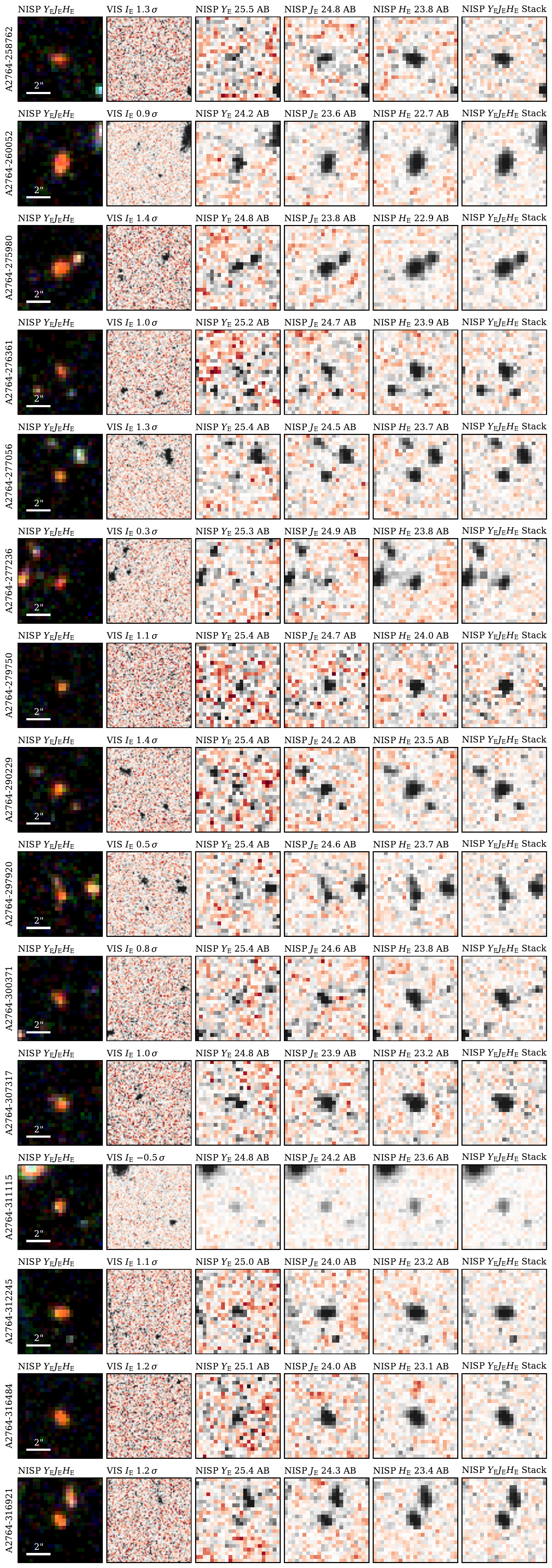}
    \end{subfigure}%
    ~ 
    \begin{subfigure}[T]{0.5\textwidth}
        \centering
        \includegraphics[width=0.92\textwidth]{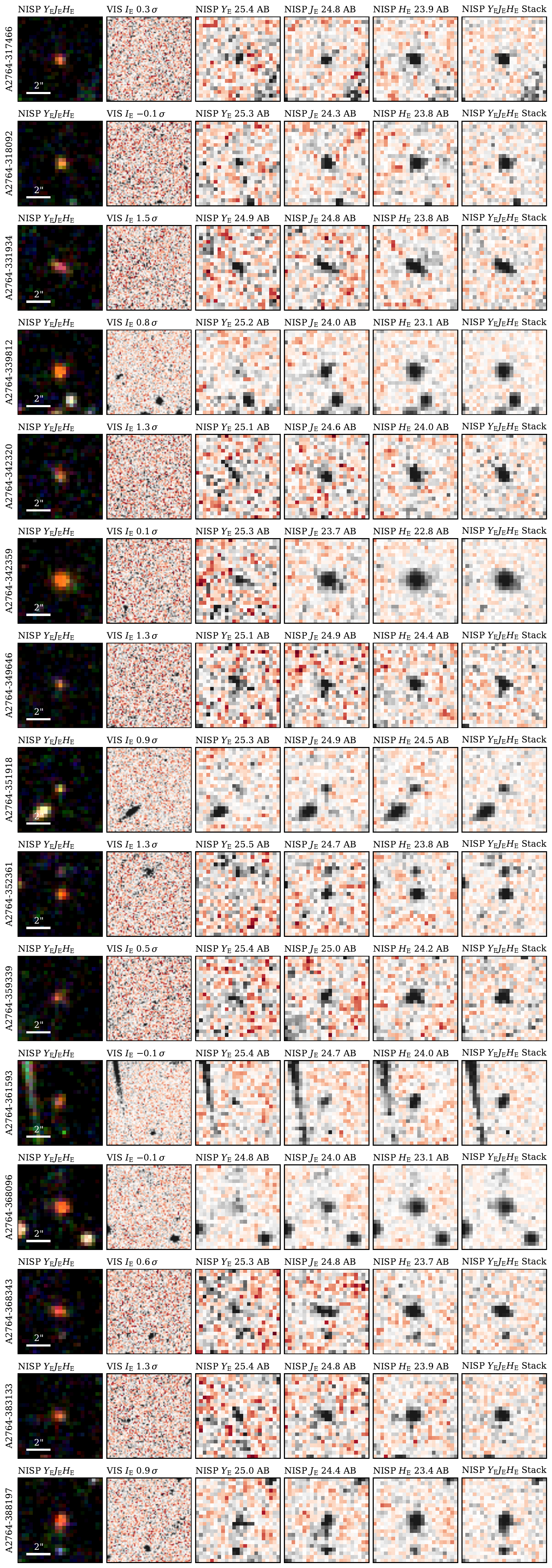}

    \end{subfigure}
    \caption{Continued...}
    \label{fig:app:ers_4}
\end{figure}
\end{appendix}

\end{document}